\documentstyle[12pt,aps,epsf,epsfig]{revtex}
\begin{document}

%*************************************************
%************** D E F I N I T I O N S ************
%**************                       ************
\def\half{{1\over 2}}
\def\be{\begin{equation}}
\def\ee{\end{equation}}
\def\3s1{$^3$S$_1$}
\def\1s0{$^1$S$_0$}
\def\3p2{$^3$P$_2$}
\def\3p1{$^3$P$_1$}
\def\3p0{$^3$P$_0$}
\def\1p1{$^1$P$_1$}

%*************************************************
\title{
\hfill \parbox[t]{2.0 in} {\rm \small
ORNL-CTP-96-03  } 
\vskip 0.1 cm
On the Mechanism of Open-Flavor Strong Decays\footnote[2]{Dedicated to
Atsje Barnes,
1 Aug 1993 - 13 Jan 1996.} 
}

\author{E.S.Ackleh,\thanks{Current address: Hewlett-Packard Corporation,
SIT-UNIX Team, M.S.906-SE3, 20 Perimeter Summit Blvd.,
Atlanta, GA 30319-1417.} T.Barnes}
\address{
Computational and Theoretical Physics Group\\
Oak Ridge National Laboratory,
Oak Ridge, TN 37831-6373, and\\
Department of Physics and Astronomy,
University of Tennessee,
Knoxville, TN 37996-1200
}

\author{E.S.Swanson}
\address{
Department of Physics,
North Carolina State University,
Raleigh, NC 27695-8202
}
\maketitle

\begin{center}
{\bf Abstract}
\end{center}

\begin{abstract}
Open-flavor strong decays 
are mediated by $q\bar q$ pair
production, which 
is known to occur dominantly with
\3p0 quantum numbers. The relation of the phenomenological
\3p0 model of these decays to ``microscopic"
QCD decay mechanisms has never been clearly established. 
In this paper we investigate $q\bar q$ meson decay amplitudes 
assuming
pair production 
from the scalar confining interaction
(sKs)
and from one gluon exchange (OGE). 
sKs pair production 
predicts 
decay amplitudes of approximately the correct magnitude and 
D/S amplitude ratios 
in
$b_1\to\omega\pi$
and
$a_1\to\rho\pi$
which are close to experiment.
The OGE
decay amplitude is found to be subdominant in most cases, a notable
exception being
$^3$P$_0\to{}^1$S$_0+{}^1$S$_0$.
The full sKs~+~OGE
amplitudes 
differ significantly from \3p0 model predictions 
in some channels and can be distinguished
experimentally, for example through an accurate comparison of
the D/S amplitude ratios in
$b_1\to\omega\pi$ and
$a_1\to\rho\pi$. 
\end{abstract}

\newpage

\section{Introduction: the $^3$P$_0$ Model.}

\subsection{Summary of the $^3$P$_0$ model.}
Strong decays constitute a rather poorly understood area of hadron
physics. A
phenomenological model
of hadron decays was developed in the
1970s by LeYaouanc {\it et al} \cite{LY1,LY2,LY3},
which assumes as earlier
suggested by Micu \cite{Micu} that during
a hadron decay a
$q\bar q$ pair is produced from the vacuum with vacuum quantum
numbers, $J^{PC} = 0^{++}$. Since this corresponds to a \3p0 
$q\bar q$ state, this is now generally referred to as the \3p0
decay model. The \3p0 pair production Hamiltonian
for the decay of a $q\bar q$ meson A to mesons B + C is usually written
in a rather complicated form with explicit wavefunctions 
\cite{3p0revs,GS,KI,GI},
which in the conventions of Geiger and Swanson \cite{GS} (to within an irrelevant
overall phase) is
\begin{displaymath}
\langle BC| H_I | A \rangle = \gamma \;
\int \!\!\! \int { d^{\, 3} r d^{\, 3} y \over (2\pi )^{3/2} } \; 
e^{{i\over 2} \vec P_B \cdot \vec r }\;
\Psi_A (\vec r)
\; 
\langle  \vec \sigma \rangle_{q\bar q}  \cdot \Big(
i\vec \nabla_B +
i\vec \nabla_C +
 \vec P_B \Big)\;
\Psi_B^* ({\vec r \over 2}  + \vec y) \; \Psi_C^*({\vec r \over 2}  - \vec y)
\end{displaymath}
\begin{equation}
\hskip 10cm \delta( \vec P_A - \vec P_B - \vec P_C \, )
\end{equation}
for all quark and antiquark masses equal.
The strength $\gamma$ of the decay interaction is
regarded as a free constant 
and is fitted to data \cite{gammanote}.

Studies of hadron decays using this model
have been
concerned almost exclusively with numerical predictions, and have not led to
any fundamental modifications.
Recent studies have considered changes
in the spatial
dependence of the pair production amplitude as a function of
quark coordinates \cite{3p0revs,GS,KI,GI}
but the fundamental decay mechanism is usually not
addressed; this is widely believed to be a nonperturbative process,
involving ``flux tube breaking".
There have been some studies of the decay
mechanism which consider an alternative phenomenological
model in which the $q\bar q$ pair is produced with $^3$S$_1$
quantum numbers \cite{ABCKP}; this possibility however is 
rejected by experiment
\cite{GS}. 
In this paper we first review the phenomenological \3p0 model
and its predictions and develop a diagrammatic representation,
following which we will test candidate 
QCD (quark-gluon) decay mechanisms for these strong decays.

An equivalent formulation of the
$^3$P$_0$ model regards the decays as due to
an interaction Hamiltonian involving Dirac quark fields,
\begin{equation}
H_I = g \, \int d^{\, 3} x \; \bar\psi \psi
\ ,
\end{equation}
which in the nonrelativistic limit gives matrix elements
identical to (1) with the identification
\begin{equation}
\gamma
 = {g \over   2  m_q}
\end{equation}
where $m_q$ is the quark mass of the produced pair.
The operator $g\bar\psi \psi$ leads to the decay
$(q\bar q)_A\to
(q\bar q)_B+
(q\bar q)_C$
through the $b^\dagger d^\dagger$ term.

This model makes no reference
to color, which if included would simply change the definition
of the interaction strength $\gamma$; since $\gamma$
is fitted to data, this would not change the
predictions for meson decays. 
An {\it ad hoc} feature of the $^3$P$_0$ model
is to allow only diagrams in which the $q\bar q$
pair separate into different final hadrons. This
was originally motivated by experiment
(specifically the weakness of $\phi\to\rho\pi$); in the
QCD-based current-current 
decay models that we shall discuss subsequently the absence of
these hairpin diagrams 
is a natural consequence of the
production of the $q\bar q$ pair in a color octet state.

To determine a decay rate we evaluate
the matrix element of the decay
Hamiltonian, which is of the form
\begin{equation}
\langle BC| H_I | A \rangle = \; h_{fi} \;
\delta( \vec A - \vec B - \vec C ) \ .
\end{equation}
 
This $h_{fi}$ decay amplitude
can be
combined with relativistic phase
space to give the differential decay rate, which is
\begin{equation}
{d \Gamma_{A\to BC} \over d\Omega_{\ \ \ \ \ } }
 =
\, 2\pi \,  {P E_B E_C \over M_A} \, |h_{fi}|^2 
\end{equation}
where we have set $\vec A = \vec 0$ and $P = |\vec B| = |\vec C|$.
An equivalent result is quoted by Geiger and Swanson \cite{GS} as the
``actual phase space" case of their equation (20). We shall usually quote
results for the amplitude 
${\cal M}_{L_{BC}S_{BC}}$, defined by
\begin{equation}
\Gamma_{A\to BC}
 = 
2\pi \,  {P E_B E_C \over M_A}  \,
\sum_{LS}
|{\cal M}_{LS}|^2\, 
\ .
\end{equation}
For $S=0$ 
this ${\cal M}_{LS}$ is the coefficient of $Y_{LM}$ in $h_{fi}$, 
and in general is the amplitude
$\langle JM(L_{BC},S_{BC})|BC\rangle$ in the 
$BC$ final state. 

We have developed a diagrammatic technique for evaluating the $h_{fi}$
matrix element, which is discussed in detail for 
$\rho\to\pi\pi$ in the \3p0 model in Appendix A.
For $q\bar q$ meson decays there are two independent diagrams, shown below.

\begin{figure}
$$\epsfxsize=4truein\epsffile{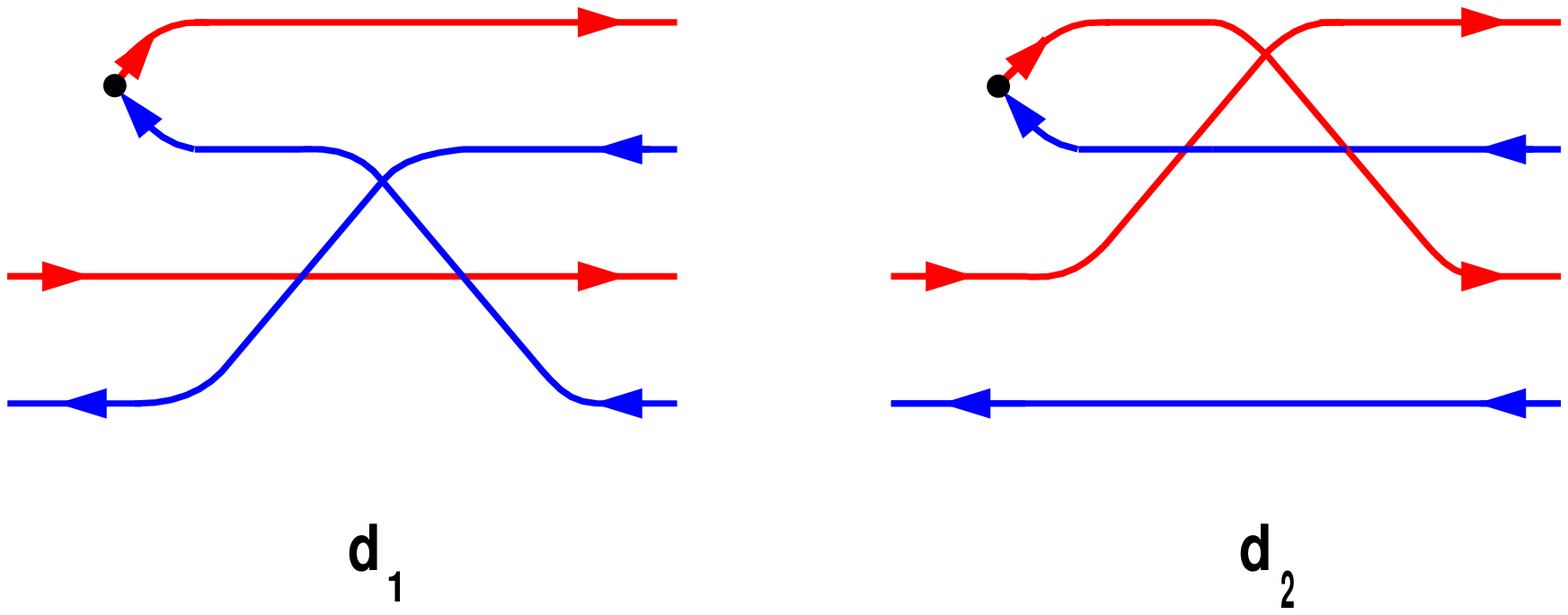}$$
{Figure~1.
The two independent $q\bar q$ meson decay diagrams in the $^3$P$_0$ decay model.
%\protect\cite{Micu}.
}
\end{figure}

\subsection{\3p0 model decay rates for light L$_{q\bar q}=0$ and
L$_{q\bar q}=1$ mesons.}

Since the \3p0
model
involves a free
phenomenological
interaction strength $\gamma$,
comparison with experiment requires
a fit to several rates (to determine both $\gamma$ and the wavefunction
parameters)
or a determination of
amplitude ratios
in decays with
more than one partial wave,
in which $\gamma$ cancels.
Here we illustrate both applications; first we will
evaluate the dominant two-body
decay rates of light nonstrange mesons with L$_{q\bar q}=0$ and
L$_{q\bar q}=1$, and
subsequently we will evaluate D/S ratios in the decays
$b_1\to \omega\pi$,
$a_1\to \rho\pi$
and
$h_1\to \rho\pi$.

An ${\cal M}_{LS}$ decay amplitude in the \3p0 model 
with SHO wavefunctions is proportional to 
a polynomial ${\cal P}_{LS}(x)$ in 
$x=P/\beta$ times an exponential,
\begin{equation}
{\cal M}_{LS} = {\gamma \over \pi^{1/4} \beta^{1/2} } \;
{\cal P}_{LS}(x) \, e^{-x^2/12} \ .
\end{equation}
For the cases considered here these polynomials are

\begin{eqnarray}
 & {\cal P}_{10}^{ ( ^3{\rm S}_1 \to ^1{\rm S}_0 + ^1{\rm S}_0 )} \  & =
-{2^5  \over 3^3 }\; x  \;
\\
 & {\cal P}_{20}^{ ( ^3{\rm P}_2 \to ^1{\rm S}_0 + ^1{\rm S}_0 )} \  & =
+{2^6  \over 3^4 5^{1/2} }\; x^2  \;
\\
 & {\cal P}_{21}^{( ^3{\rm P}_2 \to ^3{\rm S}_1 + ^1{\rm S}_0 ) }\  & =
-{2^{11/2}  \over 3^{7/2} 5^{1/2}  }\; x^2  \;
\\
 & {\cal P}_{01}^{( ^3{\rm P}_1 \to ^3{\rm S}_1 + ^1{\rm S}_0 ) }\  & =
+{2^5  \over 3^{5/2}  }\; \Big(1 - {2\over 9} x^2\Big)   \;
\\
 & {\cal P}_{21}^{( ^3{\rm P}_1 \to ^3{\rm S}_1 + ^1{\rm S}_0 ) }\  & =
-{2^{11/2}  \over 3^{9/2} }\; x^2  \;
\\
 & {\cal P}_{00}^{( ^3{\rm P}_0 \to ^1{\rm S}_0 + ^1{\rm S}_0 ) }\  & =
+{2^{9/2}  \over 3^2 }\; \Big(1 - {2\over 9} x^2\Big)  \; 
\\
 & {\cal P}_{01}^{( ^1{\rm P}_1 \to ^3{\rm S}_1 + ^1{\rm S}_0 ) }\  & =
-{2^{9/2}  \over 3^{5/2} }\;  \Big(1 - {2\over 9} x^2\Big)  \;
\\
 & {\cal P}_{21}^{( ^1{\rm P}_1 \to ^3{\rm S}_1 + ^1{\rm S}_0 ) }\  & =
-{2^6  \over 3^{9/2} }\; x^2  \; \ . 
\end{eqnarray}

For physical decays
there is an additional multiplicative flavor factor, as discussed in
Appendix~A. In figure 2 we show these decay rates for physical mesons
for a wide range of wavefunction length
scales $\beta$ (defined in App.A).

\begin{figure}
$$\epsfxsize=5truein\epsffile{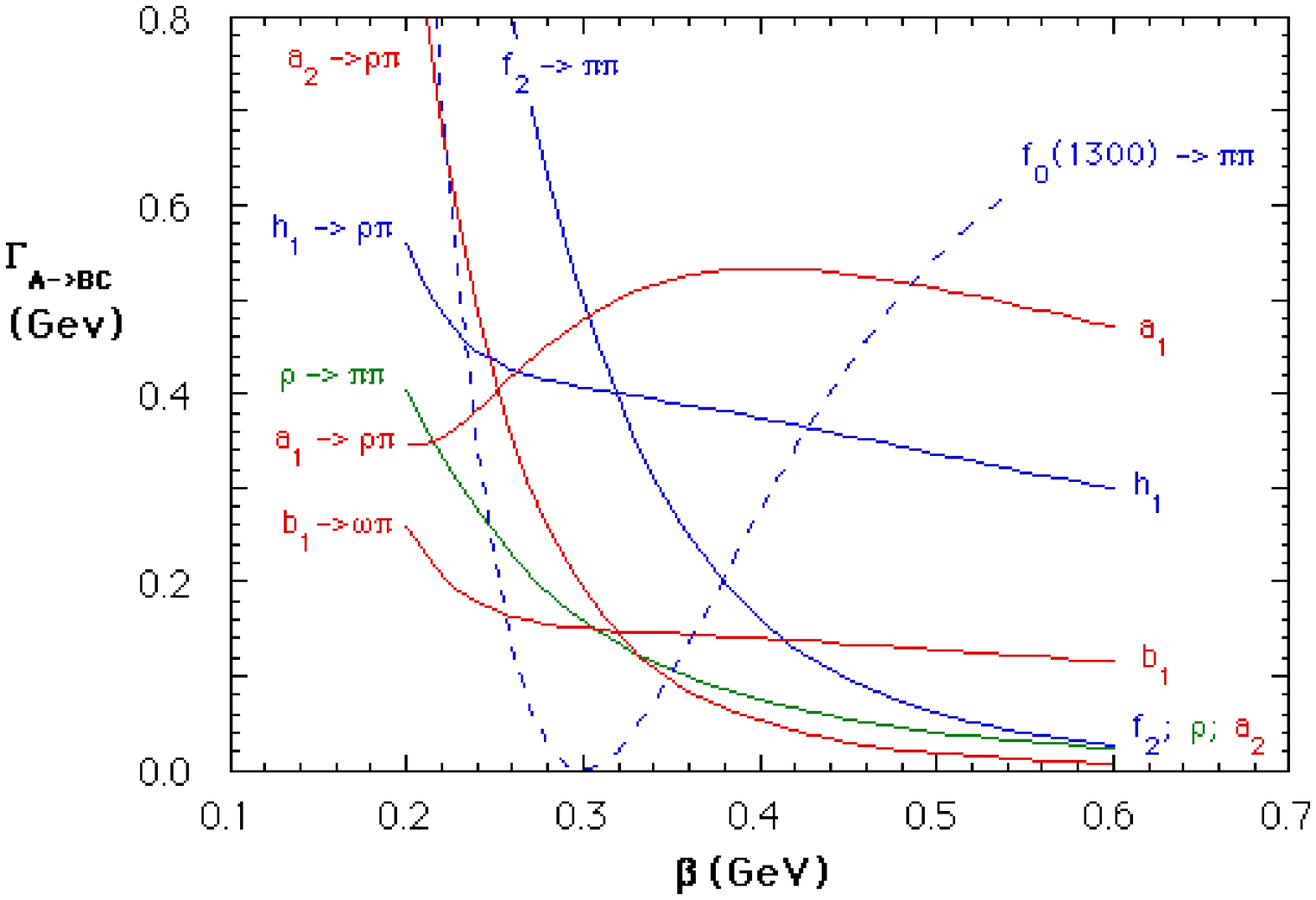}$$
{Figure~2. Representative 
light $q\bar q$ meson decay rates in the $^3$P$_0$ decay model.
Parameters $\beta$ variable, $\gamma=0.5$.
}
\end{figure}

We set $\gamma=0.5$ in the figure because this is near the
optimum fitted value.
A value of $\beta\approx 0.35$-$0.4$ GeV evidently
gives a reasonably
accurate description of the relative decay rates (see also Table 1 below). 
Many recent quark
model studies of meson \cite{3p0revs,GS,KI,GI} and baryon \cite{CR} decays 
use a value of $\beta= 0.4$ GeV, at the upper end of this range.

A simple estimate of the optimum $\beta$ 
for these SHO wavefunctions
follows if we require 
maximum overlap with
Coulomb plus linear wavefunctions $\Psi$.
With typical quark model parameters of
$\alpha_s=0.6$, $b=0.18$ GeV$^2$ and $m_q=0.33$ GeV, the overlap
$|\int \Psi^* \psi_{SHO} \; d^{\, 3}x |^2$
is maximum at
$\beta= 0.316$~GeV for L$_{q\bar q}=0$ ($99.4\% $ overlap)
and
$\beta= 0.274$~GeV for L$_{q\bar q}=1$ ($99.6\% $ overlap). 
Curiously, this $\beta$ would imply a narrow $f_0(1300)$ in the
$^3P_0$ decay model.
Comparison with figure 2 shows that these $\beta$ estimates 
are similar to the values required to describe decays,
although there is some evidence of a systematic discrepancy,
$\beta_{decay} / \beta_{Cou.+lin.} \approx 1.2$.

\vskip 1cm

\begin{center}
\begin{tabular}{||c||c|c|c|c||}  \hline
\multicolumn{5}{||c||}{Table 1. \3p0 Model Fit to Light Meson Decay Rates.} \\ \hline 
Decay  &  Expt.\cite{PDG} & \3p0 Theory &  D/S Expt \cite{PDG,ARGUS} & D/S \3p0 Theory 
\\ \hline
$\rho \to \pi\pi$ & 151. MeV & 79. MeV &     &     \\  \hline
$f_2 \to \pi\pi$ &  157. MeV & 170. MeV &     &     \\  \hline
$a_2 \to \rho\pi$ &  72. MeV & 54. MeV &     &     \\  \hline
$a_1 \to \rho\pi$ & 400. MeV & 545. MeV & $-0.09(2)$ & $-0.154$       \\  \hline
$b_1 \to \omega\pi$ & 142. MeV & 143. MeV & $+0.260(35)$ & $+0.292$       \\  \hline
$h_1 \to \rho\pi$ & 360. MeV & 383. MeV & $-$    & $-0.111$    \\  \hline
${\rm K}^*_0(1430)\to {\rm K}\pi$ & 287. MeV & 166. MeV &     &     \\  \hline
$f_0 \to \pi\pi$ & 150-400 MeV (not fitted) & 271. MeV &     &     \\  \hline
\end{tabular}
\end{center}

\vskip 1cm

In Table 1 we show the result of a fit to a representative set of
seven well established light
S- and P-wave $q\bar q$ meson decay rates, with $\beta$ and $\gamma$ 
taken as free parameters.
The masses assumed were 
$M_{\pi}=0.138$~GeV,
$M_{\rm K}=0.496$~GeV,
$M_{\rho}=0.77$~GeV,
$M_{\omega}=0.782$~GeV,
$M_{h_1}=1.17$~GeV,
$M_{a_1}=1.23$~GeV,
$M_{b_1}=1.231$~GeV,
$M_{f_0}=1.3$~GeV,
$M_{a_2}=1.318$~GeV and
$M_{{\rm K}_0^*}=1.429$~GeV.
All but ${\rm K}^*_0(1430)\to {\rm K}\pi$
are nonstrange systems, which we chose 
to avoid parameter differences due to strange quarks. The
${\rm K}^*_0(1430)$ is included because 
we found that it is quite
important to include a light \3p0 $q\bar q$ decay,
since the rate 
$ ^3$P$_0\to ^1$S$_0 + ^1$S$_0$ is quite sensitive to $\beta$, and
the ${\rm K}^*_0(1430)$ is the only well established \3p0  
resonance.
On fitting these widths using Eqs.(6-15) (minimizing
$\sum_{i=1}^7 (\Gamma_{A\to BC}^{thy.}/\Gamma_{A\to BC}^{expt.} -1 )^2$),
we find the parameters
\begin{equation}
\beta = 0.397\  {\rm GeV} 
\end{equation}
\begin{equation}
\gamma = 0.506 \ .
\end{equation}
Clearly the most important discrepancy in the table is $\rho\to\pi\pi$,
which is well known to be a problem relative to the decays of P-wave
$q\bar q$ mesons in this model. The ${\rm K}^*_0(1430)$ width is also
rather smaller than experiment. 

If we attempt constrained fits with fixed $\beta$,
we find serious disagreement
with experiment for even moderate changes 
away from $\beta=0.4$ GeV.
For example, at $\beta=0.3$ GeV we find an $f_2$ broader
than the $a_1$ and $h_1$ and a ${\rm K}^*_0(1430)$ width near zero.
Conversely, 
if we increase $\beta$
to $0.5$ GeV 
we find quite small partial widths for
$\rho\to\pi\pi$,
$a_2\to\rho\pi$ and
$f_2\to\pi\pi$ relative to the $b_1$, $a_1$ and $h_1$, 
on average only
0.31 of $b_1\to\omega\pi$; experimentally the ratio is 0.89.

The interesting question of the width of an
$f_0(1300)$
$q\bar q$ state is unfortunately
problematical in the $^3$P$_0$ model,
since like ${\rm K}^*_0(1430)\to {\rm K}\pi$ this rate
has a node near $\beta=0.30$~GeV, and increases rapidly with $\beta$
above the node. For our fitted parameters we find 
$\Gamma_{f_0(1300)\to\pi\pi} = 271$~MeV, but comparison
of the predicted and observed ${\rm K}^*_0(1430)$ widths suggests that the
actual $f_0(1300)$ partial width to $\pi\pi$ is $\approx 450$~MeV.
We shall subsequently show that 
the transition $^3$P$_0 \to ^1$S$_0 +  ^1$S$_0 $  has an unusually
large OGE decay amplitude, which may explain why the light scalars 
$f_0(1300)$ and K$^*_0(1430)$ are broad states despite the
node in the \3p0 decay amplitude.

\subsection{\3p0 results for D/S ratios}

Sensitive tests of decay models
are possible in decays with more than
one partial wave, because one can measure the relative
phases as well as the magnitudes of the decay amplitudes in these processes. 

Multiamplitude decays
require
at least one final meson to have nonzero spin. For the decays
considered here the relevant final states 
are $\rho\pi$ and $\omega\pi$; other possibilities are excluded by
phase space. These final states have S$^P=1^+$, so of all the light
L$_{q\bar q}=0$ and L$_{q\bar q}=1$ mesons only the $1^+$ mesons have
more than one
partial wave; these 
decay to both S- and D-wave vector + pseudoscalar final states.
On 
increasing L$_{q\bar q}$
we next encounter 
multiamplitude decays
in the D states, for example in
$\pi_2\to\rho\pi$ (P,F) \cite{GS} and
$\pi_2\to f_2 \pi$ (S,D,G).

Here we will calculate the D/S amplitude ratios in 
the
decays $b_1\to\omega\pi$ and $a_1\to\rho\pi$. D/S for $h_1\to\rho\pi$,
which has not been measured, is theoretically
equal to $b_1\to\omega\pi$ to within small phase space differences,
since these are
both $ ^1$P$_1 \to ^3$S$_1 + ^1$S$_0$  decays.
The $b_1$ is clearly the most attractive of these experimentally,
due to the smaller
widths of the $b_1$ and $\omega$.

We assume an initial polarization $b_1^+($J$_z=+1)$ and determine the
$h_{fi}$ matrix element to a final state with a specific $\omega$
polarization that allows both L values; we choose $\omega($S$_z=+1)$. 
Using the diagrammatic
techniques of Appendix A, we find that the two diagrams $d_1$ and $d_2$ give
equal contributions, and the total $h_{fi}$ is
\begin{equation}
h_{fi}{\bigg| }_{b_1^+(+\hat z)\to \omega(+\hat z)\pi^+} 
= - 
{\gamma \over  \pi^{1/4}
\beta^{1/2}}\;
{2^4\over  3^{5/2}}
\bigg\{
 \Big(  1 - {2\over 9} x^2\; \Big) \,
Y_{00}(\Omega)
+{2\over 3^2 5^{1/2}}x^2  \,
Y_{20}(\Omega)
\bigg\}
\; e^{- x^2/12} \ .
\end{equation}
This matrix element determines the relative S and D
amplitudes in the $|\omega\pi\rangle$ final state; since the J$^P=1^+$
$|\omega\pi\rangle $ state is
of the form
\begin{equation}
|\omega\pi \rangle = a_S | ^3{\rm S}_1 \rangle + a_D | ^3{\rm D}_1 \rangle
\end{equation}
it follows from a Clebsch-Gordon decomposition
that the amplitude to find a $(+\hat z)$-polarized $\omega$
in an $\omega\pi$ pair with decay direction $\Omega$ must be
\begin{equation}
\langle \, \omega(+\hat z)_\Omega \;
\pi_{-\Omega}\; | \, \omega\pi\, \rangle =
a_S \, Y_{00}(\Omega) + a_D \sqrt{1\over 10} \; Y_{20}(\Omega) \ .
\end{equation}
We can therefore read the $a_D/a_S$ ratio directly from the $h_{fi}$
matrix element (18), which gives
\begin{equation}
{a_D\over a_S}\bigg|_{b_1\to\omega\pi} =
+{2^{3/2} \over 3^2} \;
{ x^2 \over \big( 1 - {2\over 9}\, x^2 \big) } \ .
\end{equation}
This is equivalent to the result (4.11,12) of LeYaouanc {\it et al.}
\cite{LY1} when
one specializes their result to $R_{b_1}=R_{\omega}=R_{\pi}$, which
is our $1/\beta$.
This D/S ratio is also implicit in (14,15). 
Since the strength of the pair production amplitude and some of the
momentum dependence in the overlap integrals
cancel out in this ratio, there is less systematic uncertainty 
than in the decay rates.
Proceeding similarly for $a_1\to\rho\pi$
we find
\begin{equation}
{a_D\over a_S}\bigg|_{a_1\to\rho\pi} =
-{2^{1/2} \over 3^2} \;
{ x^2 \over \big( 1 - {2\over 9}\, x^2 \big) } \ .
\end{equation}
The {\it ratio} of D/S ratios in these decays is especially interesting
in the \3p0 model
because all the dependence on the spatial wavefunctions cancels.
We then find
\begin{equation}
{
{a_D\over a_S}\bigg|_{a_1\to\rho\pi}
\over
{a_D\over a_S}\bigg|_{b_1\to\omega\pi} } = -{1\over 2} \ .
\end{equation}
Experimentally these ratios
are
\begin{equation}
{a_D\over a_S}\bigg|_{b_1\to\omega\pi} = +0.260\pm 0.035
\end{equation}
and from a recent ARGUS measurement \cite{ARGUS}
\begin{equation}
{a_D\over a_S}\bigg|_{a_1\to\rho\pi} = -0.09\pm 0.02  \ ,
\end{equation}
so the ratio is
\begin{equation}
{
{a_D\over a_S}\bigg|_{a_1\to\rho\pi}
\over
{a_D\over a_S}\bigg|_{b_1\to\omega\pi} } = -0.35 \pm 0.09 \ .
\end{equation}
Theory and experiment for these D/S ratios are compared in figure 3;
a best fit to these D/S ratios alone gives 
$\beta = 0.448$ GeV, for which (D/S)$_{b_1\to\omega\pi}=+0.219$ and 
(D/S)$_{a_1\to\rho\pi}=-0.115$. 
(The small theoretical departure from $-1/2$ is due to phase space
differences.)

The experimental 
ratio (26) is approximately consistent with the \3p0 prediction 
but shows the need for
a more accurate experimental determination, especially in $a_1\to\rho\pi$.
This important D/S ratio 
has only been measured in one experiment; 
an improved value
would allow an interesting
test of the \3p0 model, since the OGE decay mechanism predicts a departure of
the ratio (23) from $-1/2$; we will discuss this in the next section.

In summary, the decay rates of light
L$_{q\bar q}=0$ and L$_{q\bar q}=1$
mesons (figure~2)
support $\beta\approx 0.40$ GeV, whereas the D/S amplitude ratios
in $b_1\to \omega\pi$ and $a_1\to\rho\pi$
(figure~3) suggest a larger value, $\beta \approx 0.45$ GeV.
These are both significantly larger than the $\beta\approx 0.3$ GeV which
gives the maximum overlap with Coulomb plus linear wavefunctions. 

\begin{figure}
$$\epsfxsize=6truein\epsffile{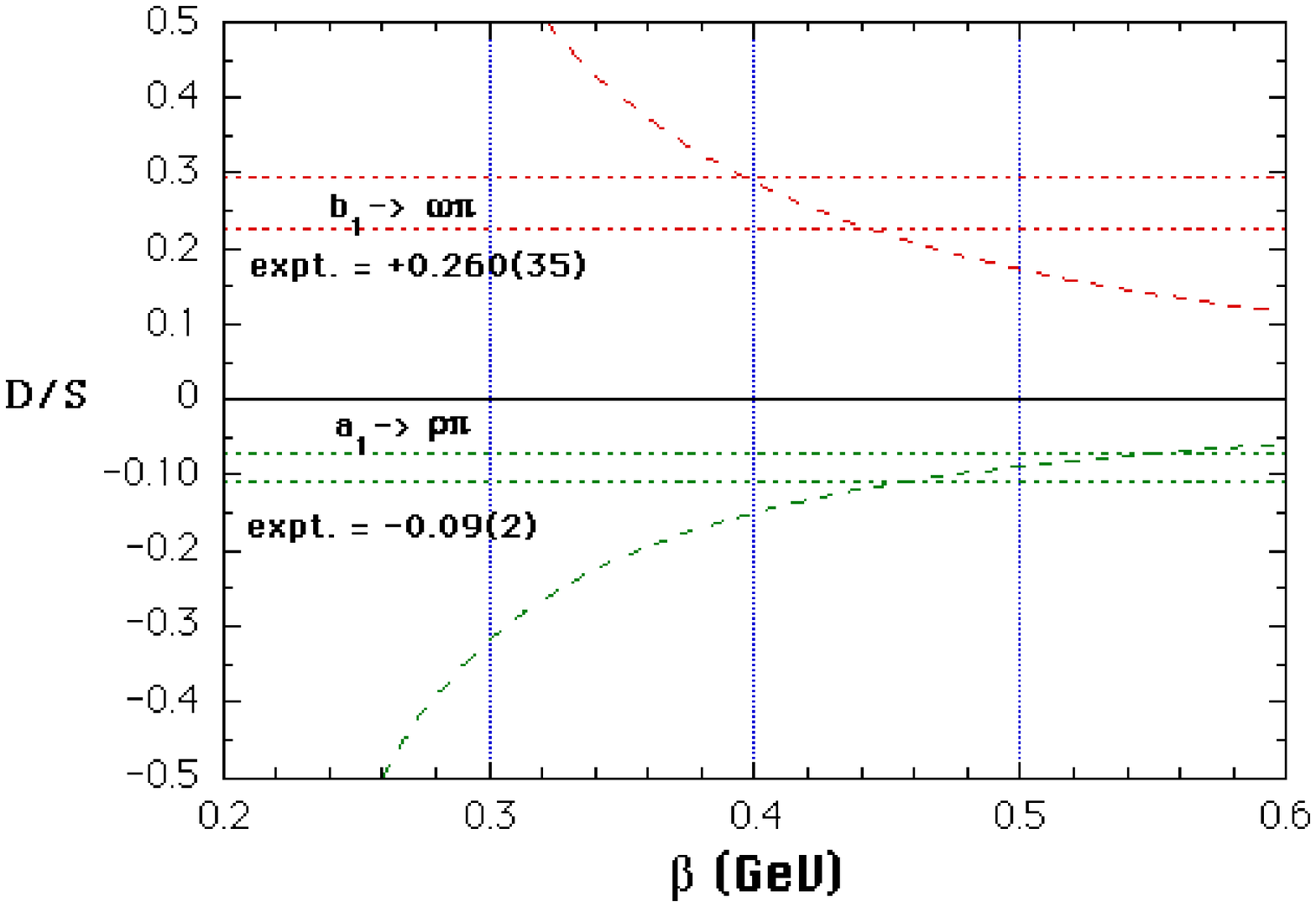}$$
{Figure~3. D/S amplitude ratios in $b_1\to \omega \pi$ 
and $a_1\to\rho\pi$ with \3p0 model 
predictions.
}
\end{figure}

\section{``Microscopic" Decay Models}

\subsection{Candidate pair production Hamiltonians.}

In microscopic decay models one attempts to describe hadron
strong decays in terms of quark and gluon degrees of freedom. 
The quark-gluon decay mechanism should give 
similar predictions to the
reasonably accurate \3p0 model,
and 
should determine the strength $\gamma$ of the \3p0 interaction 
in terms of
fundamental QCD parameters.
There has been little previous work in this area. One 
exception is the study of
open charm decays of 
$c\bar c$ 
resonances by Eichten {\it et al.} \cite{Eich}, who
assumed that decays are due to pair production from
the confining interaction. 
We will confirm this assumption
for light quarkonia in most but not all cases.
Eichten {\it et al.} however assumed
that confinement is a Lorentz {\it vector} interaction; this is now
believed to be incorrect, and the currently
accepted scalar form leads to quite different
matrix elements.

We begin by assuming that these strong decays are driven by
the 
same interquark Hamiltonian which determines the spectrum,
and that it incorporates 
scalar confinement and
one gluon exchange. 
These interactions and their associated 
decay amplitudes are undoubtedly all present and should
be added coherently. 
We will determine
decay rates for the two-body decays discussed
in the previous section,
assuming the appropriate field-theoretic generalization of the
usual interquark Hamiltonian.

The current-current interactions due to the scalar confining
interaction and OGE can be written in the generic form
\begin{equation}
H_I = {1\over 2} \int\!\!\int d^{\, 3}x\, d^{\, 3} y
\ J^a(\vec x \, ) \cdot
K(|\vec x - \vec y \, |\, ) \cdot
J^a(\vec y \, )  \ .
\end{equation}

The current $J^a$ in (27) is assumed to be a color octet.
The currents $J$ (with the color dependence
$\lambda^a/2$ factored out) and the
kernels $K(r)$ for the interactions
are 
\begin{equation}
J =
\left\{
\begin{array}{ll}
 \bar \psi \, \psi
\ &   {\rm scalar\ confining\ interaction }\\
 \psi^\dagger \psi
\ &   {\rm color \ Coulomb \ OGE }\\
( \bar \psi \vec \gamma \, \psi )_{\rm T}
\ &   {\rm transverse \ OGE }\\
\end{array}
\right.
\end{equation}

\begin{equation}
K =
\left\{
\begin{array}{ll}
+{3\over 4} \,\Big( b r + S_0 \Big) 
\ &    {\rm scalar\ confining\ interaction }   \\
+\alpha_s / r
\ &    {\rm color \ Coulomb \ OGE }  \\
 - \alpha_s / r
\ &    {\rm transverse \ OGE }  \ . \\
\end{array}
\right.
\end{equation}
The scalar confinement kernel is normalized so that
$br+S_0$ is the static scalar 
potential of a color-singlet $q\bar q$
pair.

We refer to this general type of interaction
as a JKJ decay
model, and to the specific cases considered here as
sKs, j$^0$Kj$^0$ and j$^{\rm T}$Kj$^{\rm T}$ interactions.
The decay Hamiltonian assumed by Eichten {\it et al.} \cite{Eich}
in their Eq. (3.2) is a special case of our form (27) with a
j$^0$Kj$^0$ interaction.

\subsection{JKJ A$\to$BC matrix elements: general results.}

The A$\to$BC decay matrix element of the
JKJ
Hamiltonian (27)
involves a pair-production current matrix element
$\langle q\bar q | J |0\rangle $
times a scattering matrix element
$\langle q_f | J |q_i \rangle $. Diagrammatically this corresponds
to an interaction between an initial line and the produced pair,
and there are four such diagrams (see figure 4).
We label these using the $^3$P$_0$ quark line diagram labels $d_1$ and
$d_2$,
with an additional $q$ or $\bar q$ subscript denoting which
initial line the produced pair interacts with.

We specialize to the diagram $d_{1q}$ for illustration. As in the
$^3$P$_0$ model (Appendix A) there is a Fermi ``signature" phase,
a flavor factor, and a spin+space overlap integral, 

\begin{equation}
\langle BC| H_I | A \rangle_{d_{1q}} =
I_{\rm signature} \cdot
I_{\rm color} \cdot
I_{\rm flavor} \cdot
{\bf I}_{\rm spin+space} 
\end{equation}
where
\begin{equation}
{\bf I}_{spin+space} = 
I_{spin+space} \; \delta(\vec A - \vec B - \vec C \; ) \ .
\end{equation}

\begin{figure}
$$\epsfxsize=4truein\epsffile{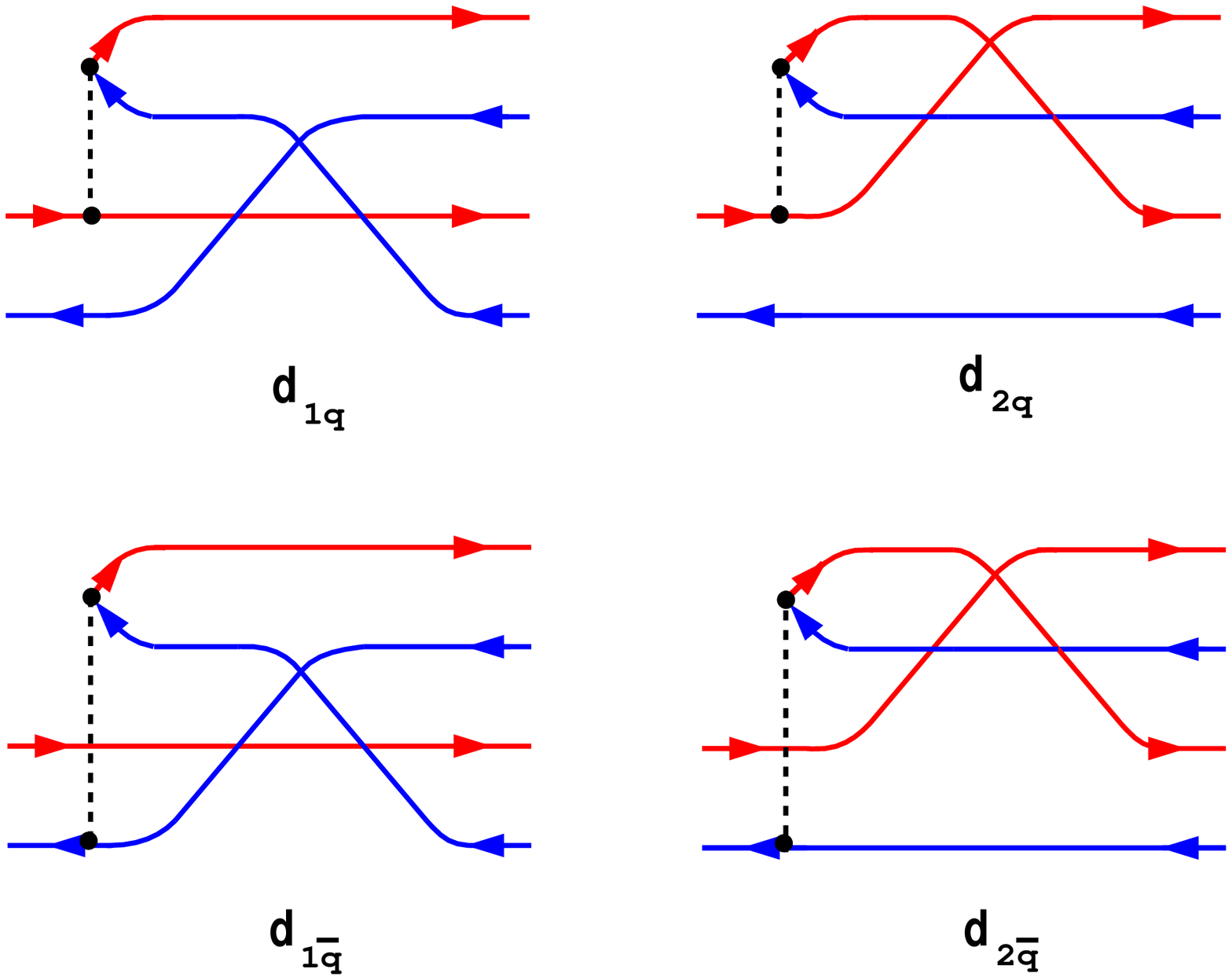}$$
{Figure~4.
The four independent $q\bar q$ meson decay diagrams in JKJ decay models.
}
\end{figure}

The signature is again $I_{\rm signature}=(-1)^3$ because there are three
line crossings. 
A new
feature due to our explicit treatment of color is a
color matrix element;
the color factor for this diagram is
\begin{equation}
I_{\rm color} = {1\over 3^{3/2}}\;
{\rm Tr}
\Big\{ {\lambda^a\over 2}\, {\lambda^a\over 2} \Big\} = +{2^2\over 3^{3/2}}
\end{equation}
and is the same for diagram $d_{2q}$ and opposite for 
$d_{1\bar q}$ and
$d_{2\bar q}$.
The flavor and spin factors are specific to the reaction and will be
discussed subsequently.
The $d_{1q}$ spatial overlap integral analogous to (A5) is
\begin{displaymath}
{\bf I}_{space} (d_{1q}) =
 2!  \int\!\!\int \! d^{\, 3}x\, d^{\, 3}y
\; {1\over 2}
K(|\vec x - \vec y \, |\, )
\int \!\!  \int \!\!  \int \!
d^{\, 3}a\, d^{\, 3}b \, d^{\, 3}c \;
\phi_A(2\vec a - \vec A ) \,
\phi_B^*(2\vec b - \vec B ) \,
\phi_C^*(2\vec c - \vec C ) \,
\end{displaymath}
\begin{equation}
\delta(\vec A - \vec a - \vec B + \vec b \, ) \;
\langle c |  J(\vec x \, ) | a \rangle \;
\langle b\bar c |  J(\vec y \, ) | 0 \rangle \ ,
\end{equation}
where the matrix elements of the currents (with
$\Gamma = \gamma^0, \gamma^i, $I) are
\begin{equation}
\langle q' | J(\vec x \, ) | q \rangle \,
= +{1\over (2\pi )^3 } {m_q \over \sqrt{ E_q E_{q'} } } \;
e^{i(q'-q)\cdot x} \;
\Big[ \bar u_{q' s'} \Gamma u_{q s} \Big]
\end{equation}
\begin{equation}
\langle \bar q{\, '} | J(\vec x \, ) | \bar q \rangle \,
= -{1\over (2\pi )^3 } {m_q \over \sqrt{ E_{\bar q} E_{\bar q'} } } \;
e^{i({\bar q'} - \bar q)\cdot x} \;
\Big[ \bar v_{\bar q \bar s} \Gamma v_{\bar q' \bar s'} \Big]
\end{equation}
and
\begin{equation}
\langle q\bar q |  J(\vec y \, ) | 0 \rangle \,
= {1\over (2\pi )^3 } {m_q \over \sqrt{ E_q E_{\bar q} } } \;
e^{i(q+\bar q)\cdot y} \;
\Big[ \bar u_{q s} \Gamma v_{{\bar q} {\bar s}}\Big]  \ .
\end{equation}
With these substitutions, and introducing a Fourier transformed kernel
\begin{equation}
{\cal K}(Q)
=  \int \! d^{\, 3}x \, e^{i\vec Q \cdot \vec r\, } K( r \, )
\end{equation}
we find
\begin{displaymath}
I_{space} (d_{1q}) =
{1\over (2\pi )^3 }
\int \!\!  \int \!
d^{\, 3}a\,  d^{\, 3}c \;
\phi_A(2\vec a - \vec A ) \;
\phi_B^*(2\vec a - 2\vec C - \vec B ) \;
\phi_C^*(2\vec c - \vec C )
\end{displaymath}
\begin{equation}
{m_q^2 \over \sqrt{ E_a E_b E_c E_{\bar c} } }
\Big[ \bar u_{b s_b} \Gamma v_{{\bar c} s_{\bar c}} \Big]
\; {\cal K}(|\vec a - \vec c \, |) \;
\Big[ \bar u_{c s_c} \Gamma u_{a s_a} \Big] 
\end{equation}
and there are
implicit momentum constraints
$\vec b = \vec a - \vec C$ and
$\vec {\bar c} = \vec C - \vec c$ from the spatial integrations in (33) and
the meson wavefunctions.

As in the $^3$P$_0$ model we take the nonrelativistic limit of this
interaction for our decay amplitude, although it would be interesting
in future work
to investigate the effect of keeping the full relativistic amplitude (38).
We also specialize to the rest frame, $\vec A = 0$ and $\vec B = -\vec C$.
These substitutions give our final general result for the $d_{1q}$ overlap
integral,
\begin{displaymath}
I_{space} (d_{1q}) =
{1\over (2\pi )^3 }
\int \!\!  \int \!
d^{\, 3}a\,  d^{\, 3}c \;
\phi_A(2\vec a) \;
\phi_B^*(2\vec a + \vec B ) \;
\phi_C^*(2\vec c + \vec B )
\end{displaymath}
\begin{equation}
\Big[ \bar u_{b s_b} \Gamma v_{{\bar c} s_{\bar c}} \Big]
\; {\cal K}(|\vec a - \vec c \, |) \;
\Big[ \bar u_{c s_c} \Gamma u_{a s_a} \Big] \ .
\end{equation}
The overlap integrals associated with the three remaining diagrams are
given in Appendix~B, and the detailed evaluation of a decay amplitude
${\cal M}_{LS}$ and a decay rate is illustrated in Appendix~C
for the decay $\rho\to\pi\pi$ with a j$^0$Kj$^0$ interaction.

\subsection{JKJ A$\to$BC matrix elements: explicit results for SHO wavefunctions.}

The complete decay amplitudes for the channels spanned by
our representative set of light two-body meson
decays are given below.
These include 
decay amplitudes from 1) linear scalar, 2) constant scalar, 
3) Coulomb OGE and 4) transverse OGE terms; a common factor is
removed,
\begin{equation}
{\cal M}_{LS} = 
{\beta^{1/2}  \over  \pi^{3/4} m_q }   \; e^{-x^2/12} \cdot
\tilde{\cal M}_{LS} \ ,
\end{equation} 
and the transverse OGE amplitude is shown as separate 4) $\delta_{ij}$ and
5) $-Q_i Q_j / \vec Q^2 $ contributions.
These five results for each channel are

\begin{eqnarray}
%
% reaction: 3S1 > 1S0 1S0
  &\tilde{\cal M}_{10}^{ ( ^3{\rm S}_1 \to ^1{\rm S}_0 + ^1{\rm S}_0 )}  =  &
\nonumber\\
&   &   \nonumber\\
% sKs
  &     &-{2^3 5\over 3^3}
         \bigg( {b\over  \beta^2 }\bigg) 
        \, x \, 
\bigg[ 
{}_1{\rm F}_1\Big(-{1\over 2};{3\over 2};\xi \Big)
+ {4\over 45}
{}_1{\rm F}_1\Big(-{1\over 2};{5\over 2};\xi \Big)
\bigg]  
\nonumber\\
% S0 
 &    &-{2^9 \over 3^{11/2} } \pi^{1/2}
       \bigg( {S_0 \over  \beta  }\bigg) 
       \, x \,
\nonumber\\
% j0Kj0
  &   &+{2^3 \over 3^3} 
       \alpha_s   
       \, x \,
\bigg[
{}_1{\rm F}_1\Big({1\over 2};{3\over 2};\xi \Big)
- {2\over 3}
{}_1{\rm F}_1\Big({1\over 2};{5\over 2};\xi \Big)
\bigg] 
\nonumber\\
% jKj d_ij
 &    &+{2^3 \over 3^3}
       \alpha_s 
       \, x \,
\bigg[ 
{}_1{\rm F}_1\Big({1\over 2};{3\over 2};\xi \Big)
+ {10\over 9}
{}_1{\rm F}_1\Big({1\over 2};{5\over 2};\xi \Big)
\bigg]    
\nonumber\\
% jKj -q_iq_j/q^2 
 &    &-{2^3 \over 3^2}
       \alpha_s 
       \, x \,
\bigg[ 
{}_1{\rm F}_1\Big({1\over 2};{3\over 2};\xi \Big)
- {2\over 3}
{}_1{\rm F}_1\Big({1\over 2};{5\over 2};\xi \Big)
\bigg]  \ ,  
\\
&   &   \nonumber\\
 &\tilde{\cal M}_{20}^{ ( ^3{\rm P}_2 \to ^1{\rm S}_0 + ^1{\rm S}_0 )} = &
\nonumber\\
&   &   \nonumber\\
% sKs
 &       & +{2^2\; 5^{1/2} \over 3^3}
               \bigg( {b\over  \beta^2 }\bigg) 
               \, x^2 \, 
\bigg[ 
{}_1{\rm F}_1\Big(-{1\over 2};{3\over 2};\xi \Big)
+ {8\over 15}
{}_1{\rm F}_1\Big(-{1\over 2};{5\over 2};\xi \Big)
+ {8\over 225}
{}_1{\rm F}_1\Big(-{1\over 2};{7\over 2};\xi \Big)
\bigg] 
\nonumber\\
% S0 
   &  &+{2^{10} \over 3^{13/2} 5^{1/2} } \pi^{1/2}
       \bigg( {S_0 \over  \beta  }\bigg) 
       \, x^2 \, 
\nonumber\\
% j0Kj0
 &    &-{2^2 \over 3^3 5^{1/2} } 
       \alpha_s   
       \, x^2 \,
\bigg[
{}_1{\rm F}_1\Big({1\over 2};{3\over 2};\xi \Big)
- {4\over 9}
{}_1{\rm F}_1\Big({1\over 2};{5\over 2};\xi \Big)
- {8\over 45}
{}_1{\rm F}_1\Big({1\over 2};{7\over 2};\xi \Big)
\bigg] 
\nonumber\\
% jKj d_ij
  &   &-{2^2 \over 3^3 5^{1/2} }
       \alpha_s 
       \, x^2 \,
\bigg[ 
{}_1{\rm F}_1\Big({1\over 2};{3\over 2};\xi \Big)
+ {4 \over 3}
{}_1{\rm F}_1\Big({1\over 2};{5\over 2};\xi \Big)
+ {8 \over 27}
{}_1{\rm F}_1\Big({1\over 2};{7\over 2};\xi \Big)
\bigg]   
\nonumber\\
%
% jKj -q_iq_j/q^2 
%
 &    &+{2^2  \over 3^2 5^{1/2} }
       \alpha_s 
       \, x^2 \,
\bigg[ 
{}_1{\rm F}_1\Big({1\over 2};{3\over 2};\xi \Big)
- {8 \over 27}
{}_1{\rm F}_1\Big({1\over 2};{5\over 2};\xi \Big)
- {8 \over 27}
{}_1{\rm F}_1\Big({1\over 2};{7\over 2};\xi \Big)
\bigg]  \ ,  
\\
&   &   \nonumber\\
 & \tilde{\cal M}_{21}^{ ( ^3{\rm P}_2 \to ^3{\rm S}_1 + ^1{\rm S}_0 )}= & 
\nonumber\\
&   &   \nonumber\\
% sKs
 &     &-{2^{3/2} 5^{1/2}\over 3^{5/2}}
         \bigg( {b\over  \beta^2 }\bigg) 
        \, x^2 \, 
\bigg[ 
{}_1{\rm F}_1\Big(-{1\over 2};{3\over 2};\xi \Big)
+ {8\over 15}
{}_1{\rm F}_1\Big(-{1\over 2};{5\over 2};\xi \Big)
+ {8\over 225}
{}_1{\rm F}_1\Big(-{1\over 2};{7\over 2};\xi \Big)
\bigg] 
\nonumber\\
% S0 
 &    &-{2^{19/2} \over 3^6 5^{1/2} } \pi^{1/2}
       \bigg( {S_0 \over  \beta  }\bigg) 
       \, x^2 \,
\nonumber\\
% j0Kj0
 &    &+{2^{3/2} \over 3^{5/2} 5^{1/2} } 
       \alpha_s   
       \, x^2 \,
\bigg[
{}_1{\rm F}_1\Big({1\over 2};{3\over 2};\xi \Big)
- {4\over 9}
{}_1{\rm F}_1\Big({1\over 2};{5\over 2};\xi \Big)
- {8\over 45}
{}_1{\rm F}_1\Big({1\over 2};{7\over 2};\xi \Big)
\bigg]
\nonumber\\
% jKj d_ij
 &     &+{2^{5/2} \over 3^{5/2} 5^{1/2} }
       \alpha_s 
       \, x^2 \,
\bigg[ 
{}_1{\rm F}_1\Big({1\over 2};{3\over 2};\xi \Big)
+ {4\over 9}
{}_1{\rm F}_1\Big({1\over 2};{5\over 2};\xi \Big)
+ {8\over 135}
{}_1{\rm F}_1\Big({1\over 2};{7\over 2};\xi \Big)
\bigg] \  
\nonumber\\
%
% jKj -q_iq_j/q^2 
%
 &    &-{2^{3/2}  \over 3^{3/2} 5^{1/2} }
       \alpha_s 
       \, x^2 \,
\bigg[ 
{}_1{\rm F}_1\Big({1\over 2};{3\over 2};\xi \Big)
- {8 \over 27}
{}_1{\rm F}_1\Big({1\over 2};{5\over 2};\xi \Big)
- {8 \over 27}
{}_1{\rm F}_1\Big({1\over 2};{7\over 2};\xi \Big)
\bigg]  \ ,  
\\
& & \nonumber\\
 & \tilde{\cal M}_{01}^{ ( ^3{\rm P}_1 \to ^3{\rm S}_1 + ^1{\rm S}_0 )} = &
\nonumber\\
&   &   \nonumber\\
% sKs
  &     &-{2^55\over 3^{7/2}}
         \bigg( {b\over  \beta^2 }\bigg) 
        \,  
\bigg[ 
{}_1{\rm F}_1\Big(-{3\over 2};-{1\over 2};\xi \Big)
- {12 \over 5}
{}_1{\rm F}_1\Big(-{3\over 2};{1\over 2};\xi \Big)
+ {8\over 15}
{}_1{\rm F}_1\Big(-{3\over 2};{3\over 2};\xi \Big)
\bigg] 
\nonumber\\
% S0 
 &    &+{2^9 \over 3^5  } \pi^{1/2}
       \bigg( {S_0 \over  \beta  }\bigg) 
      \Big( 1 - {2\over 9}x^2 \Big)
\nonumber\\
% j0Kj0
 &    &+{2^5 \over 3^{5/2}  } 
       \alpha_s   
        \,
\bigg[
{}_1{\rm F}_1\Big(-{1\over 2};-{1\over 2};\xi \Big)
+ {4\over 3}
{}_1{\rm F}_1\Big(-{1\over 2};{1\over 2};\xi \Big)
- {8\over 3}
{}_1{\rm F}_1\Big(-{1\over 2};{3\over 2};\xi \Big)
\bigg]
\nonumber\\
% jKj d_ij
 &     &+{2^6 \over 3^{5/2} }
       \alpha_s 
       \, 
\bigg[ 
{}_1{\rm F}_1\Big(-{1\over 2};-{1\over 2};\xi \Big)
- {10 \over 3}
{}_1{\rm F}_1\Big(-{1\over 2};{1\over 2};\xi \Big)
+ {28\over 9}
{}_1{\rm F}_1\Big(-{1\over 2};{3\over 2};\xi \Big)
\bigg]    
\nonumber\\
%
% jKj -q_iq_j/q^2 
%
 &    &-{2^5 \over 3^{3/2} }
       \alpha_s 
       \, 
\bigg[ 
{}_1{\rm F}_1\Big(-{1\over 2};-{1\over 2};\xi \Big)
+ {8 \over 9}
{}_1{\rm F}_1\Big(-{1\over 2};{1\over 2};\xi \Big)
- {16 \over 9}
{}_1{\rm F}_1\Big(-{1\over 2};{3\over 2};\xi \Big)
\bigg]  \ ,  
\\
& & \nonumber\\
 & \tilde{\cal M}_{21}^{ ( ^3{\rm P}_1 \to ^3{\rm S}_1 + ^1{\rm S}_0 )} = &
\nonumber\\
&   &   \nonumber\\
% sKs
  &     &-{2^{3/2}5\over 3^{7/2}}
         \bigg( {b\over  \beta^2 }\bigg) 
        \, x^2 \, 
\bigg[ 
{}_1{\rm F}_1\Big(-{1\over 2};{3\over 2};\xi \Big)
+ {8\over 15} 
{}_1{\rm F}_1\Big(-{1\over 2};{5\over 2};\xi \Big)
+ {8\over 225} 
{}_1{\rm F}_1\Big(-{1\over 2};{7\over 2};\xi \Big)
\bigg] 
\nonumber\\
% S0
 &    &-{2^{19/2} \over 3^7  } \pi^{1/2}
       \bigg( {S_0 \over  \beta  }\bigg)
       \, x^2
\nonumber\\
% j0Kj0
 &    &+{2^{3/2} \over 3^{7/2}  }           
       \alpha_s  
        \, x^2 \,
\bigg[
{}_1{\rm F}_1\Big({1\over 2};{3\over 2};\xi \Big)
- {4\over 9}
{}_1{\rm F}_1\Big({1\over 2};{5\over 2};\xi \Big)
- {8\over 45} 
{}_1{\rm F}_1\Big({1\over 2};{7\over 2};\xi \Big)
\bigg]
\nonumber\\
% jKj d_ij
 &     &+{2^{5/2} \over 3^{7/2} }
       \alpha_s
       \, x^2 \, 
\bigg[
{}_1{\rm F}_1\Big({1\over 2};{3\over 2};\xi \Big)
+ {4\over 9}
{}_1{\rm F}_1\Big({1\over 2};{5\over 2};\xi \Big)
+ {8\over 135}
{}_1{\rm F}_1\Big({1\over 2};{7\over 2};\xi \Big)
\bigg]    
\nonumber\\
%
%
% jKj -q_iq_j/q^2 
%
 &    &-{2^{3/2}  \over  3^{5/2}}
       \alpha_s 
       \, x^2 \,
\bigg[ 
{}_1{\rm F}_1\Big({1\over 2};{3\over 2};\xi \Big)
- {8 \over 27}
{}_1{\rm F}_1\Big({1\over 2};{5\over 2};\xi \Big)
- {8 \over 27}
{}_1{\rm F}_1\Big({1\over 2};{7\over 2};\xi \Big)
\bigg]  \ ,  
\\
 & & \nonumber\\
 & \tilde{\cal M}_{00}^{ ( ^3{\rm P}_0 \to ^1{\rm S}_0 + ^1{\rm S}_0 )} = &
\nonumber\\
 &   &   \nonumber\\
% sKs
  &     &-{2^{9/2}5\over 3^3}
         \bigg( {b\over  \beta^2 }\bigg) 
        \,  
\bigg[ 
{}_1{\rm F}_1\Big(-{3\over 2};-{1\over 2};\xi \Big)
- {12 \over 5}
{}_1{\rm F}_1\Big(-{3\over 2};{1\over 2};\xi \Big)
+ {8\over 15}
{}_1{\rm F}_1\Big(-{3\over 2};{3\over 2};\xi \Big)
\bigg] 
\nonumber\\
% S0
 &    &+{2^{17/2} \over 3^{9/2}  } \pi^{1/2}
       \bigg( {S_0 \over  \beta  }\bigg)
      \Big( 1 - {2\over 9}x^2 \Big)
       \, 
\nonumber\\
% j0Kj0
 &    &+{2^{9/2} \over 3^2  } 
       \alpha_s   
        \,
\bigg[
{}_1{\rm F}_1\Big(-{1\over 2};-{1\over 2};\xi \Big)
+ {4\over 3}
{}_1{\rm F}_1\Big(-{1\over 2};{1\over 2};\xi \Big)
- {8\over 3}
{}_1{\rm F}_1\Big(-{1\over 2};{3\over 2};\xi \Big)
\bigg]
\nonumber\\
% jKj d_ij
 &     &+{2^{9/2} \over 3^2 }
       \alpha_s
       \, 
\bigg[
{}_1{\rm F}_1\Big(-{1\over 2};-{1\over 2};\xi \Big)
- 8\, 
{}_1{\rm F}_1\Big(-{1\over 2};{1\over 2};\xi \Big)
+ {80 \over 9} 
{}_1{\rm F}_1\Big(-{1\over 2};{3\over 2};\xi \Big)
\bigg]    
\nonumber\\
%
% jKj -q_iq_j/q^2 
%
 &    &-{2^{9/2} \over 3 }
       \alpha_s 
       \, 
\bigg[ 
{}_1{\rm F}_1\Big(-{1\over 2};-{1\over 2};\xi \Big)
+ {8 \over 9}
{}_1{\rm F}_1\Big(-{1\over 2};{1\over 2};\xi \Big)
- {16 \over 9}
{}_1{\rm F}_1\Big(-{1\over 2};{3\over 2};\xi \Big)
\bigg]  \ ,  
\\
& & \nonumber\\
 & \tilde{\cal M}_{01}^{ ( ^1{\rm P}_1 \to ^3{\rm S}_1 + ^1{\rm S}_0 )} = &
\nonumber\\
&   &   \nonumber\\
% sKs
  &     &+{2^{9/2}5\over 3^{7/2}}
         \bigg( {b\over  \beta^2 }\bigg) 
        \,  
\bigg[ 
{}_1{\rm F}_1\Big(-{3\over 2};-{1\over 2};\xi \Big)
- {12 \over 5}
{}_1{\rm F}_1\Big(-{3\over 2};{1\over 2};\xi \Big)
+ {8\over 15}
{}_1{\rm F}_1\Big(-{3\over 2};{3\over 2};\xi \Big)
\bigg] 
\nonumber\\
% S0
 &    &-{2^{17/2} \over 3^5  } \pi^{1/2}
       \bigg( {S_0 \over  \beta  }\bigg)
      \Big( 1 - {2\over 9}x^2 \Big)
       \, 
\nonumber\\
% j0Kj0
 &    &-{2^{9/2} \over 3^{5/2}  } 
       \alpha_s   
        \,
\bigg[
{}_1{\rm F}_1\Big(-{1\over 2};-{1\over 2};\xi \Big)
+ {4\over 3}
{}_1{\rm F}_1\Big(-{1\over 2};{1\over 2};\xi \Big)
- {8\over 3}
{}_1{\rm F}_1\Big(-{1\over 2};{3\over 2};\xi \Big)
\bigg]
\nonumber\\
% jKj d_ij
 &     &-{2^{9/2} \over 3^{3/2} }
       \alpha_s
       \, 
\bigg[
{}_1{\rm F}_1\Big(-{1\over 2};-{1\over 2};\xi \Big)
- {16\over 9}
{}_1{\rm F}_1\Big(-{1\over 2};{1\over 2};\xi \Big)
+ {32\over 27} 
{}_1{\rm F}_1\Big(-{1\over 2};{3\over 2};\xi \Big)
\bigg]    
\nonumber\\
%
% jKj -q_iq_j/q^2 
%
 &    &+{2^{9/2} \over 3^{3/2} }
       \alpha_s 
       \, 
\bigg[ 
{}_1{\rm F}_1\Big(-{1\over 2};-{1\over 2};\xi \Big)
+ {8 \over 9}
{}_1{\rm F}_1\Big(-{1\over 2};{1\over 2};\xi \Big)
- {16 \over 9}
{}_1{\rm F}_1\Big(-{1\over 2};{3\over 2};\xi \Big)
\bigg]  \ ,  
\\
& & \nonumber\\
 & \tilde{\cal M}_{21}^{ ( ^1{\rm P}_1 \to ^3{\rm S}_1 + ^1{\rm S}_0 )} = &
\nonumber\\
&   &   \nonumber\\
% sKs
  &     &-{2^2\; 5\over 3^{7/2}}
         \bigg( {b\over  \beta^2 }\bigg)
        \, x^2 \,
\bigg[
{}_1{\rm F}_1\Big(-{1\over 2};{3\over 2};\xi \Big)
+ {8\over 15} 
{}_1{\rm F}_1\Big(-{1\over 2};{5\over 2};\xi \Big)
+ {8\over 225} 
{}_1{\rm F}_1\Big(-{1\over 2};{7\over 2};\xi \Big)
\bigg]
\nonumber\\
% S0
 &    &-{2^{10} \over 3^7  } \pi^{1/2}
       \bigg( {S_0 \over  \beta  }\bigg)
       \, x^2 \,
\nonumber\\
% j0Kj0
 &    &+{2^2 \over 3^{7/2}  }
       \alpha_s
        \, x^2 \,
\bigg[
{}_1{\rm F}_1\Big({1\over 2};{3\over 2};\xi \Big)
- {4\over 9}
{}_1{\rm F}_1\Big({1\over 2};{5\over 2};\xi \Big)
- {8\over 45} 
{}_1{\rm F}_1\Big({1\over 2};{7\over 2};\xi \Big)
\bigg]
\nonumber\\
% jKj d_ij
 &     &+{2^2 \over 3^{5/2} }
       \alpha_s
       \, x^2 \,
\bigg[
{}_1{\rm F}_1\Big({1\over 2};{3\over 2};\xi \Big)
+ {4\over 27}
{}_1{\rm F}_1\Big({1\over 2};{5\over 2};\xi \Big)
- {8\over 405} 
{}_1{\rm F}_1\Big({1\over 2};{7\over 2};\xi \Big)
\bigg] 
\nonumber\\
%
%
% jKj -q_iq_j/q^2 
%
 &    &-{2^{2}  \over  3^{5/2}}
       \alpha_s 
       \, x^2 \,
\bigg[ 
{}_1{\rm F}_1\Big({1\over 2};{3\over 2};\xi \Big)
- {8 \over 27}
{}_1{\rm F}_1\Big({1\over 2};{5\over 2};\xi \Big)
- {8 \over 27}
{}_1{\rm F}_1\Big({1\over 2};{7\over 2};\xi \Big)
\bigg]  \ .  
\end{eqnarray}

\subsection{JKJ decay rates: numerical results.}

To estimate the numerical importance of the scalar and OGE decay 
mechanisms we first specialize to 
$\rho\to\pi\pi$ and consider each contribution in isolation.
Using (6), (40) and (41), and the flavor factors in App.A,
the decay rate due to linear scalar pair production alone would be
\begin{equation}
\Gamma_{\rho\to\pi\pi}^{\rm s K s } =
\pi^{-1/2} \Big( {2^6 5^2\over 3^6} \Big)
\Big( { b \over m_q\beta } \Big)^2
{E_\pi^2\over M_\rho }
x^3\;
\bigg[
{}_1{\rm F}_1\Big(-{1\over 2};{3\over 2};\xi \Big)
+ {4\over 45}
{}_1{\rm F}_1\Big(-{1\over 2};{5\over 2};\xi \Big)
\bigg]^2 \,
e^{-x^2 / 6 } \ .
\end{equation}
The analogous 
decay rates due to Coulomb OGE and transverse OGE 
are
\begin{equation}
\Gamma_{\rho\to\pi\pi}^{ {\rm j}^0{\rm Kj}^0 }
= \pi^{-1/2} \Big( {2^6\over 3^6} \Big)
\,  \alpha_s^2 \, \Big( {\beta \over m_q } \Big)^2
{E_\pi^2\over M_\rho }
x^3\;
\bigg[
{}_1{\rm F}_1\Big({1\over 2};{3\over 2};\xi \Big)
- {2\over 3}
{}_1{\rm F}_1\Big({1\over 2};{5\over 2};\xi \Big)
\bigg]^2 \,
e^{-x^2 / 6 } 
\end{equation}
and
\begin{equation}
\Gamma_{\rho\to\pi\pi}^{ {\rm j}^{\rm T}{\rm K j}^{\rm T}  } =
 \pi^{-1/2} \Big( {2^8\over 3^6} \Big)
\, \alpha_s^2 \, \Big( {\beta \over m_q } \Big)^2
{E_\pi^2\over M_\rho }
x^3\;
\bigg[
{}_1{\rm F}_1\Big({1\over 2};{3\over 2};\xi \Big)
+ {4 \over 9}
{}_1{\rm F}_1\Big({1\over 2};{5\over 2};\xi \Big)
\bigg]^2 \,
e^{-x^2 / 6 } \ .
\end{equation}
These rates
are shown in figure~5 
for the parameter set
$\alpha_s=0.6$, $b=0.18$ GeV$^2$ and $m_q=0.33$ GeV. 
Evidently the dominant decay mechanism is the sKs
interaction, 
pair production through the scalar confining potential.
At $\beta=0.4$ GeV the $\rho\to\pi\pi$ width predicted by the sKs model alone
is 330 MeV, about twice the experimental 151 MeV.
In comparison, transverse OGE gives a width of 3.9 MeV, and the color
Coulomb interaction gives only 0.36 MeV. (The transverse and Coulomb OGE
contributions add constructively, so the total width from OGE alone 
would be 6.7 MeV. These both interfere destructively with the 
dominant sKs amplitude,
so the total width we find from all three amplitudes is 
$\Gamma_{\rho\to\pi\pi} = 243$ MeV.)

\begin{figure}
$$\epsfxsize=6truein\epsffile{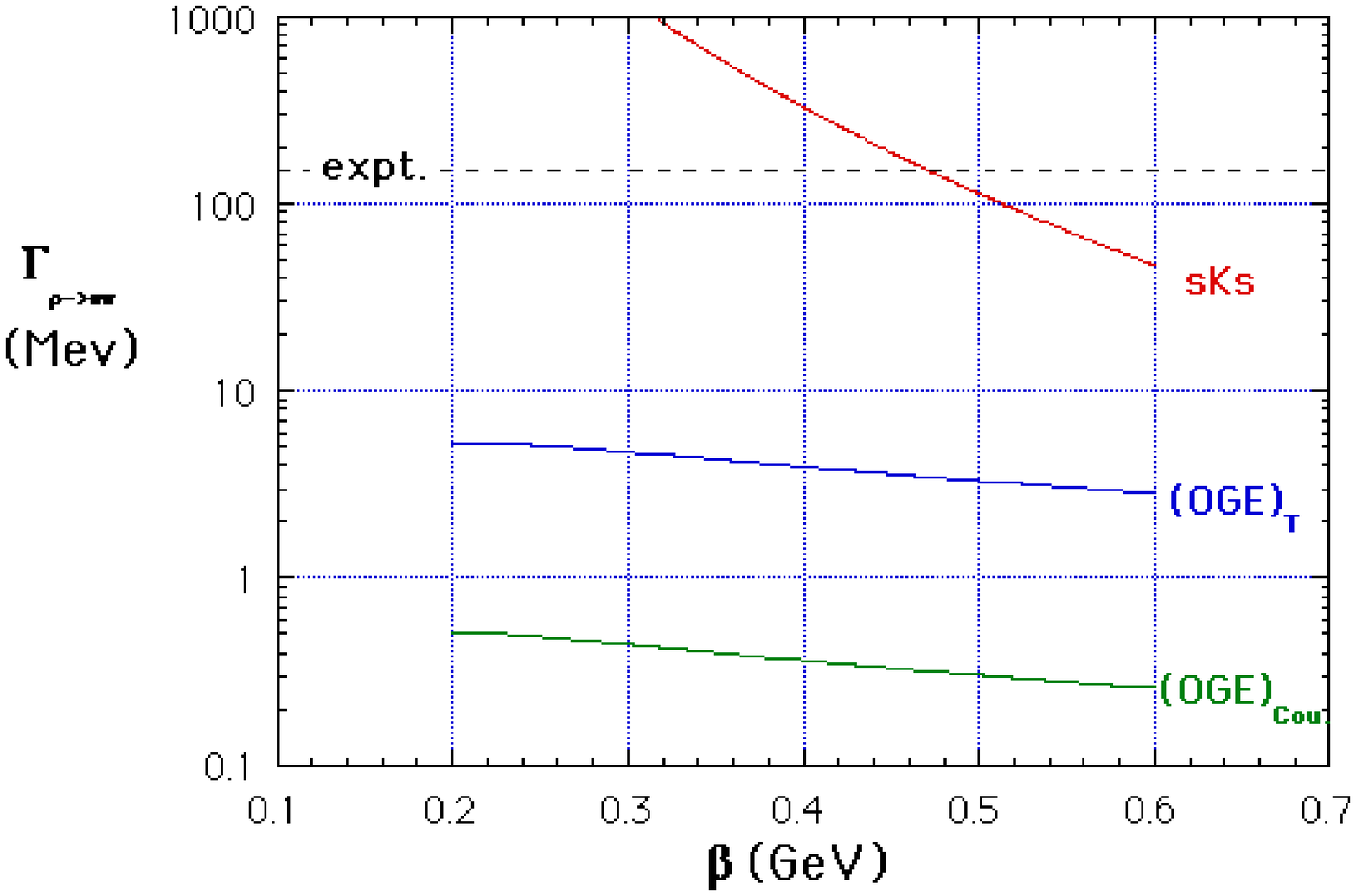}$$
{Figure~5.
The decay rate $\Gamma_{\rho\to\pi\pi}$ assuming only sKs, 
j$^{\rm T}$Kj$^{\rm T}$
or j$^0$Kj$^0$ decay interactions. Parameters $\alpha_s=0.6$, $b=0.18$ GeV$^2$, 
$m_q=0.33$ GeV.
}
\end{figure}

Although the sKs interaction is usually found to be dominant in 
our representative set of decays,
the OGE contributions 
are often 
comparable to sKs
and cannot generally be ignored. To illustrate this, in figure~6 
we show the numerical
decay amplitudes we find in each decay.
The normalization incorporates phase space, so these amplitudes 
squared give the physical decay rates.
The
magnitude of the experimental amplitude is indicated by horizontal 
lines. (The $h_1$ amplitudes are not shown because they are essentially
identical to the $b_1$ amplitudes times a flavor factor of $\sqrt{3}$.)

\hbox to \hsize{%
\begin{minipage}[t]{0.5\hsize}
\begin{figure}
\epsfxsize=2.9in
\hbox to \hsize{\hss\epsffile{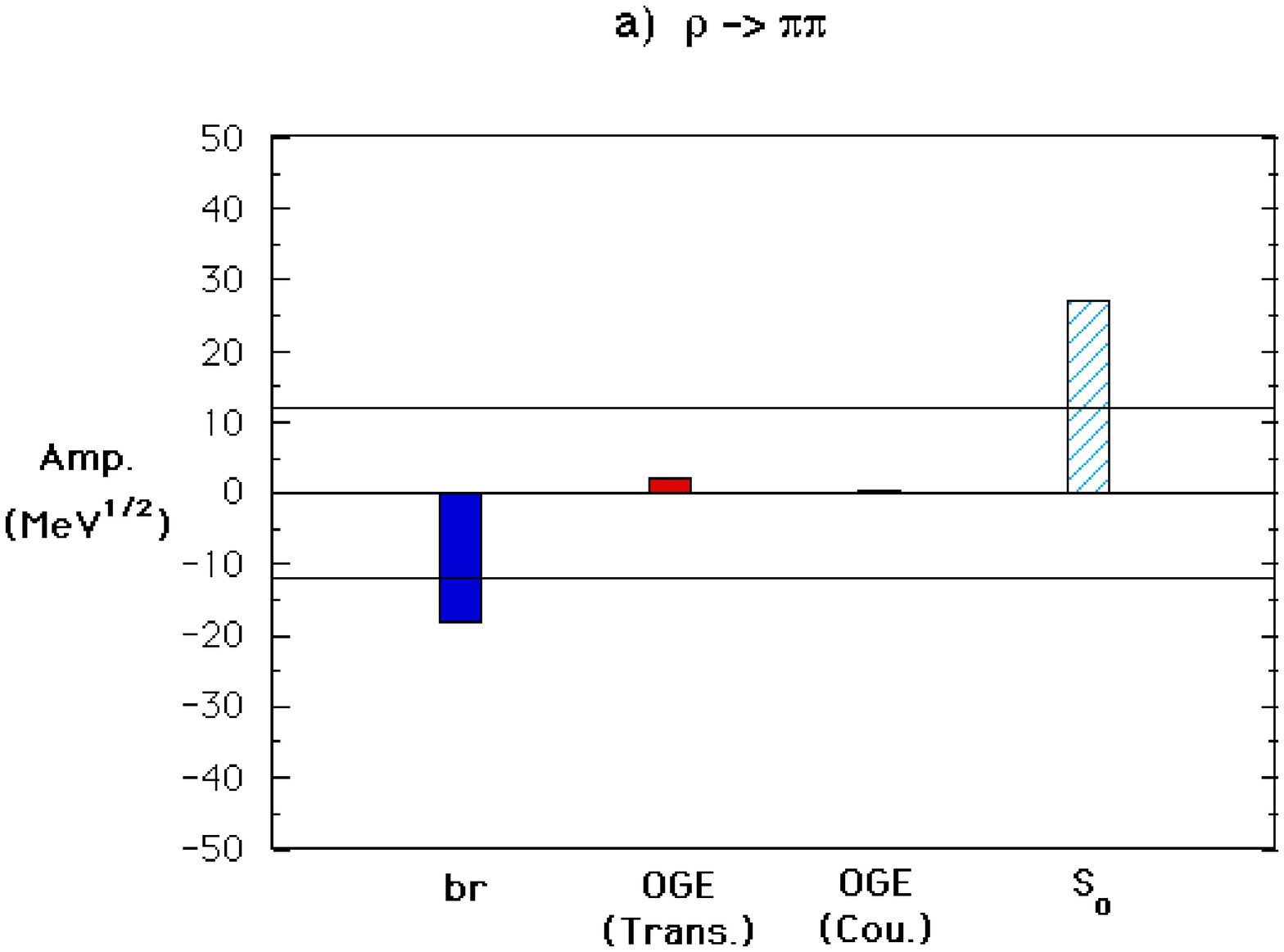}\hss}
\end{figure}
\end{minipage}
\begin{minipage}[t]{0.5\hsize}
\begin{figure}
\epsfxsize=2.9in
\hbox to \hsize{\hss \epsffile{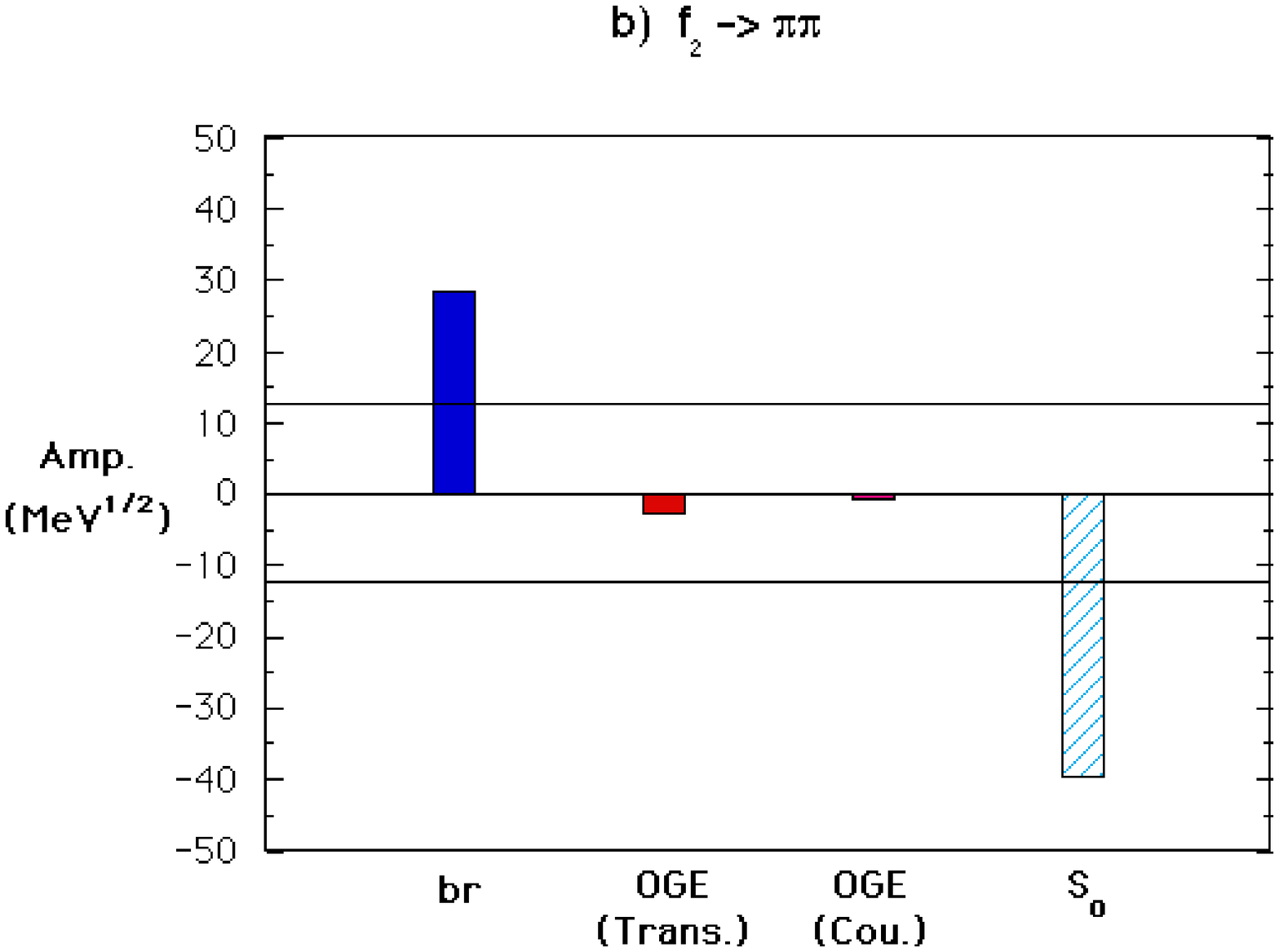}\hss}
\end{figure}
\end{minipage}}

\hbox to \hsize{%
\begin{minipage}[t]{0.5\hsize}
\begin{figure}
\epsfxsize=2.9in
\hbox to \hsize{\hss\epsffile{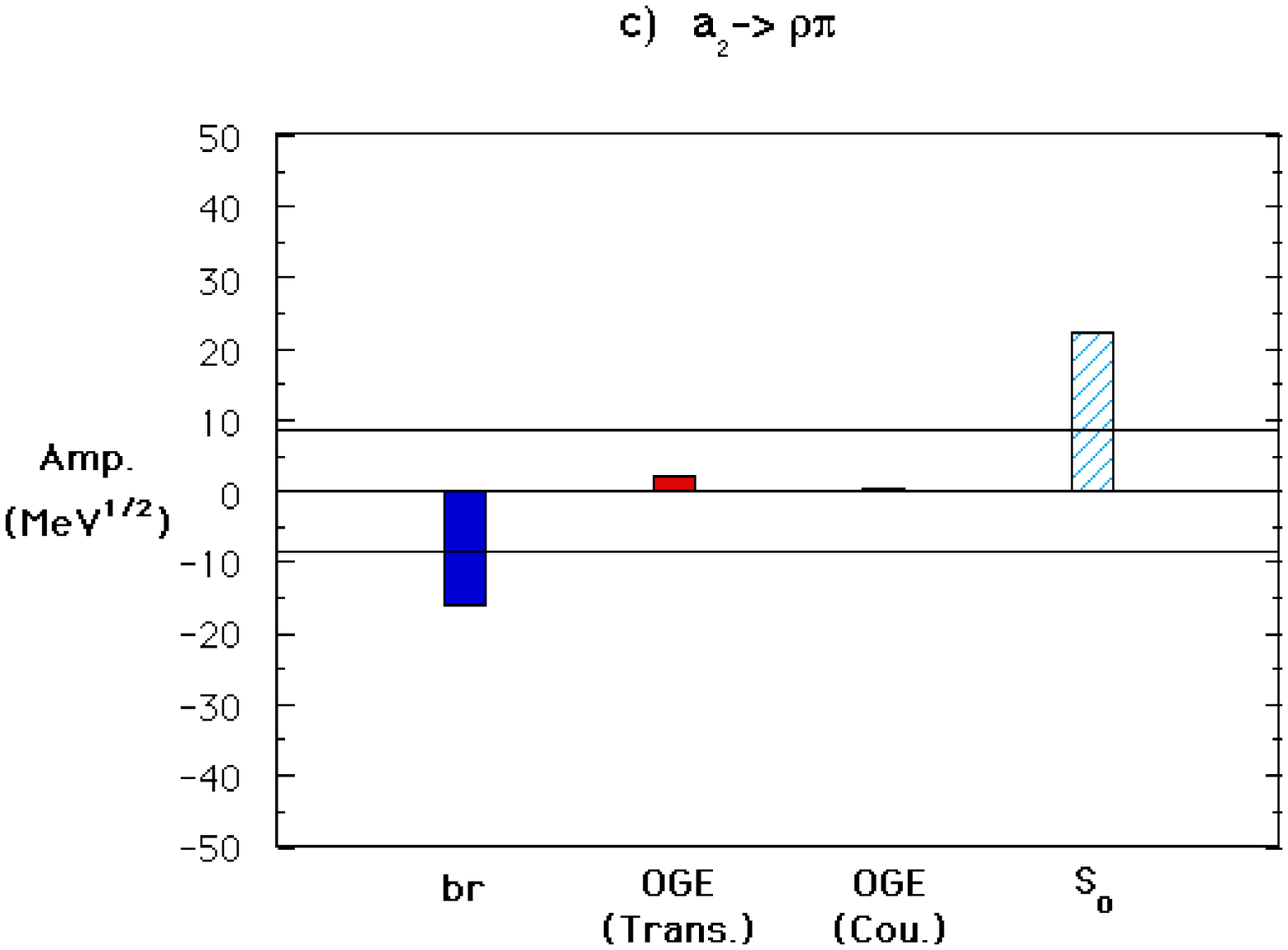}\hss}
\end{figure}
\end{minipage}
}

%\begin{minipage}[t]{0.5\hsize}
%\begin{figure}
%\epsfxsize=2.9in
%\hbox to \hsize{\hss \epsffile{}\hss}
%\end{figure}
%\end{minipage}}

\hbox to \hsize{%
\begin{minipage}[t]{0.5\hsize}
\begin{figure}
\epsfxsize=2.9in
\hbox to \hsize{\hss\epsffile{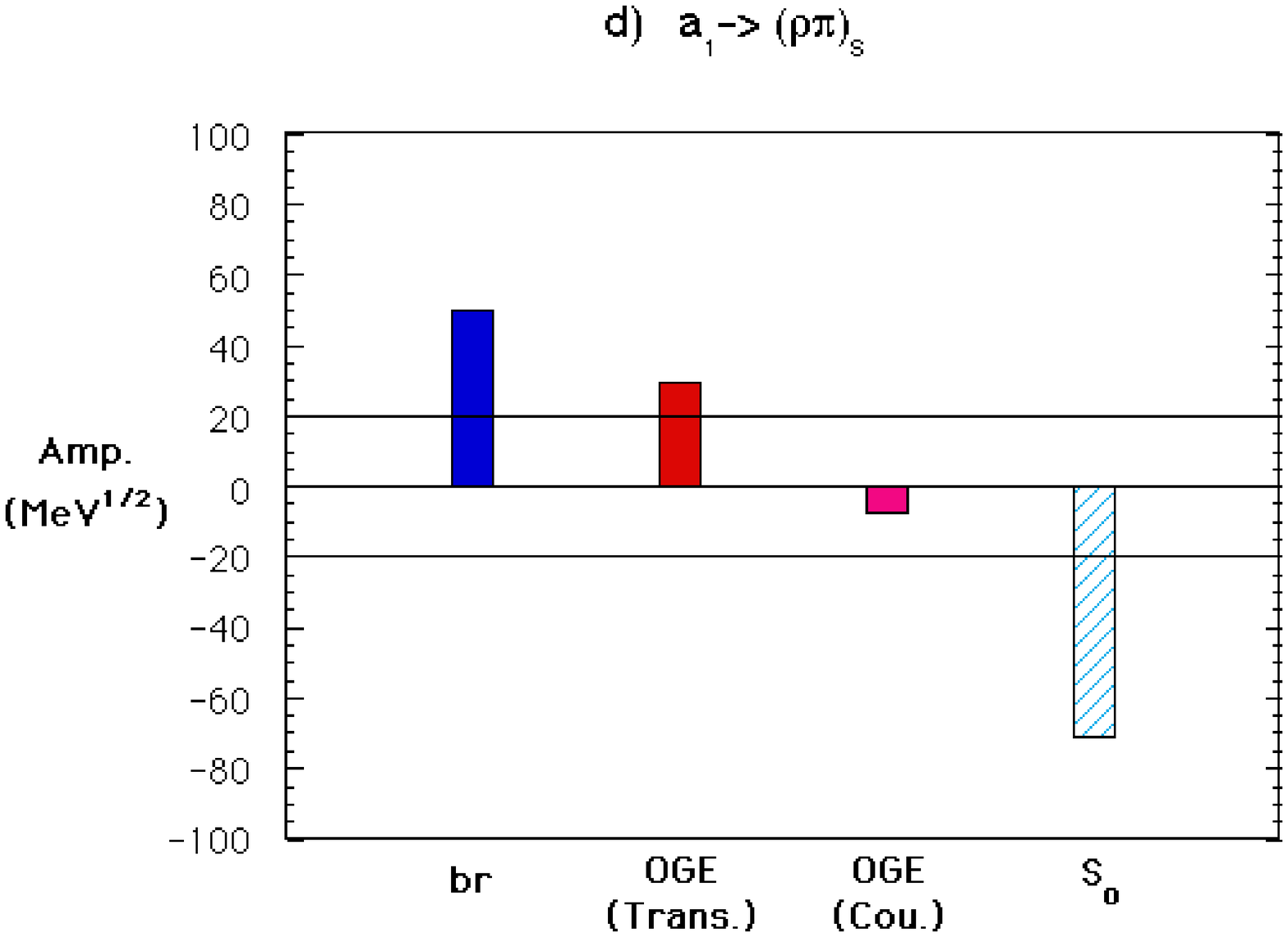}\hss}
\end{figure}
\end{minipage}
\begin{minipage}[t]{0.5\hsize}
\begin{figure}
\epsfxsize=2.9in
\hbox to \hsize{\hss \epsffile{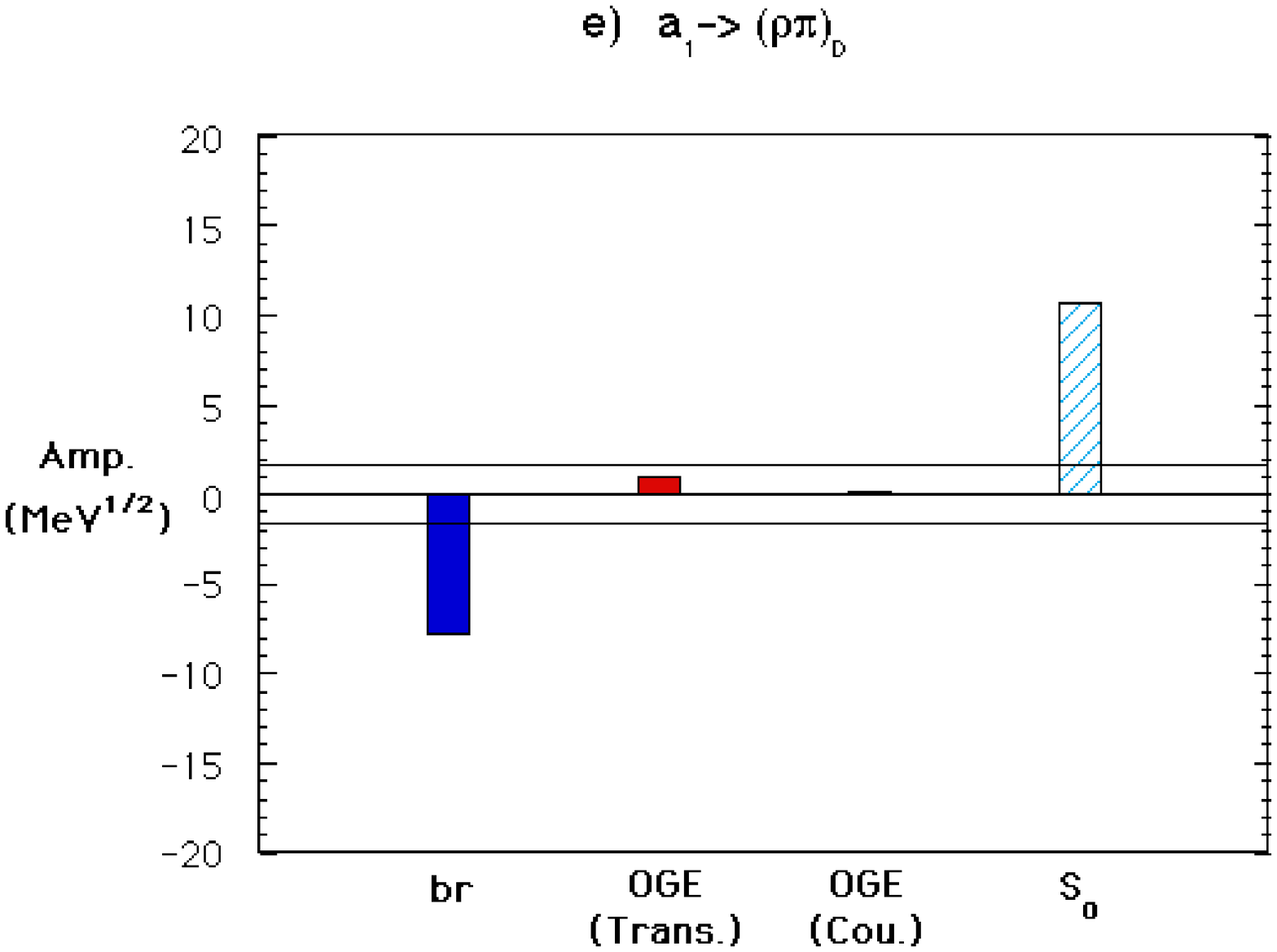}\hss}
\end{figure}
\end{minipage}}

\hbox to \hsize{%
\begin{minipage}[t]{0.5\hsize}
\begin{figure}
\epsfxsize=2.9in
\hbox to \hsize{\hss\epsffile{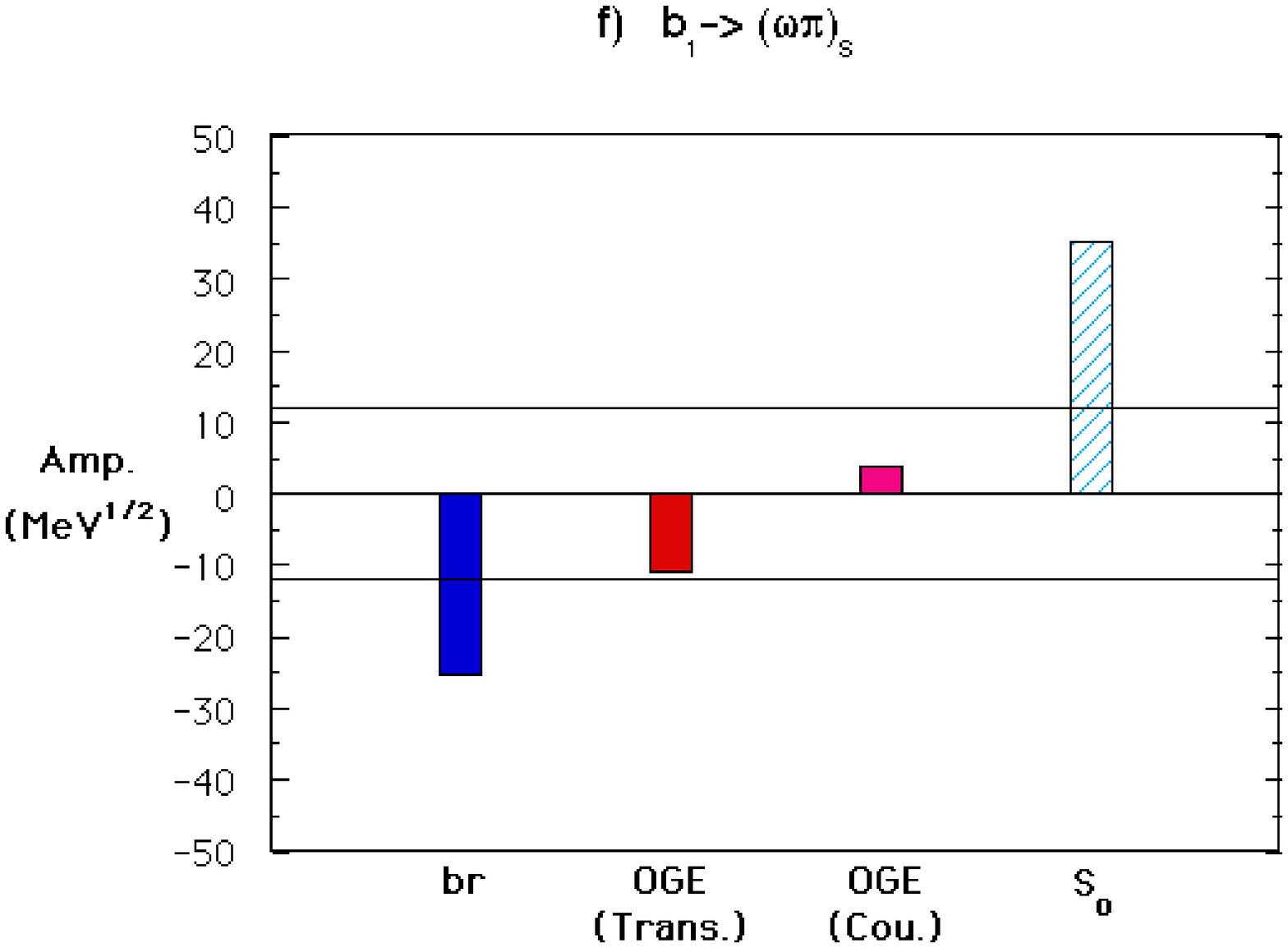}\hss}
\end{figure}
\end{minipage}
\begin{minipage}[t]{0.5\hsize}
\begin{figure}
\epsfxsize=2.9in
\hbox to \hsize{\hss \epsffile{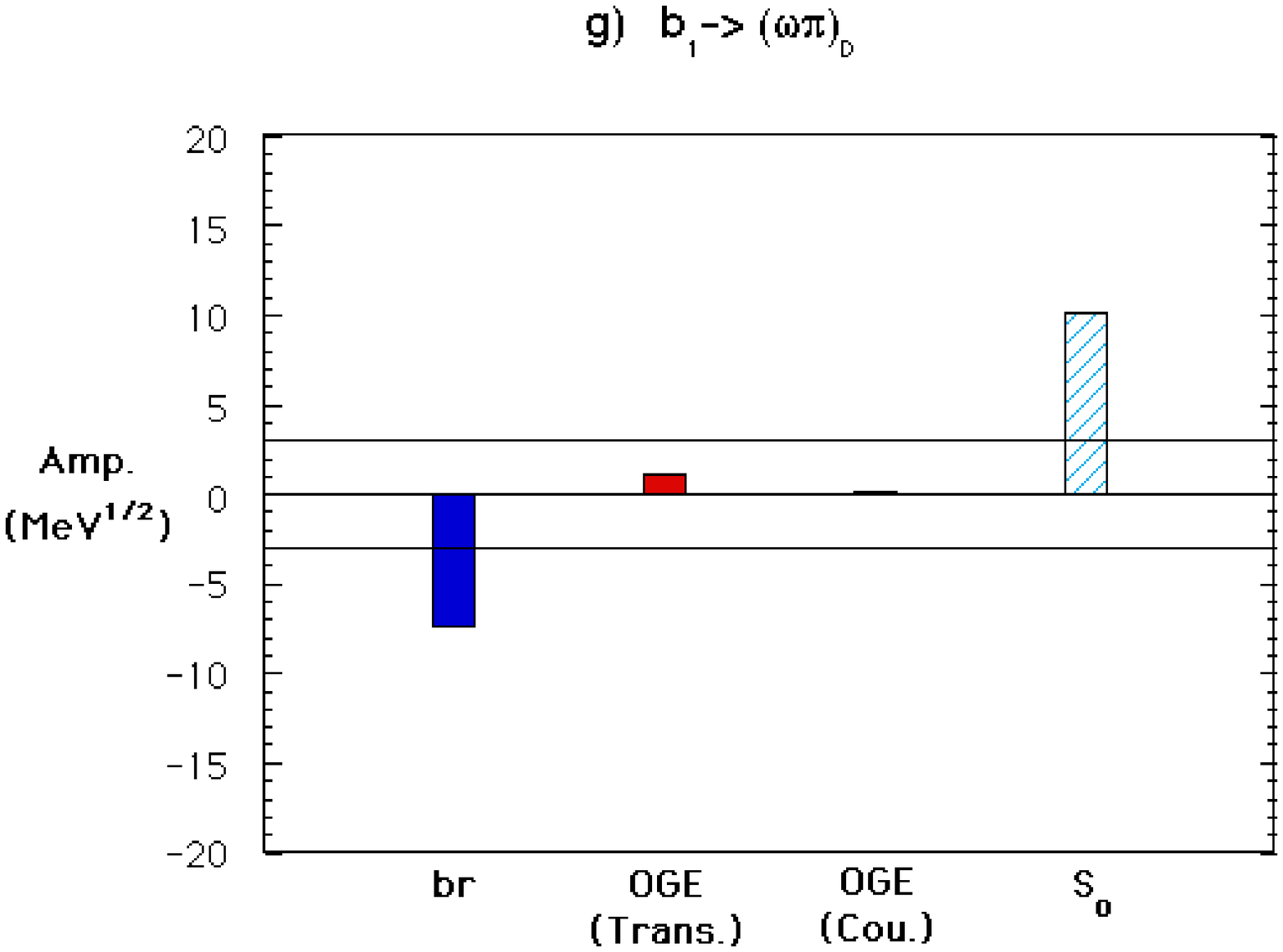}\hss}
\end{figure}
\end{minipage}}

\hbox to \hsize{%
\begin{minipage}[t]{0.5\hsize}
\begin{figure}
\epsfxsize=2.9in
\hbox to \hsize{\hss\epsffile{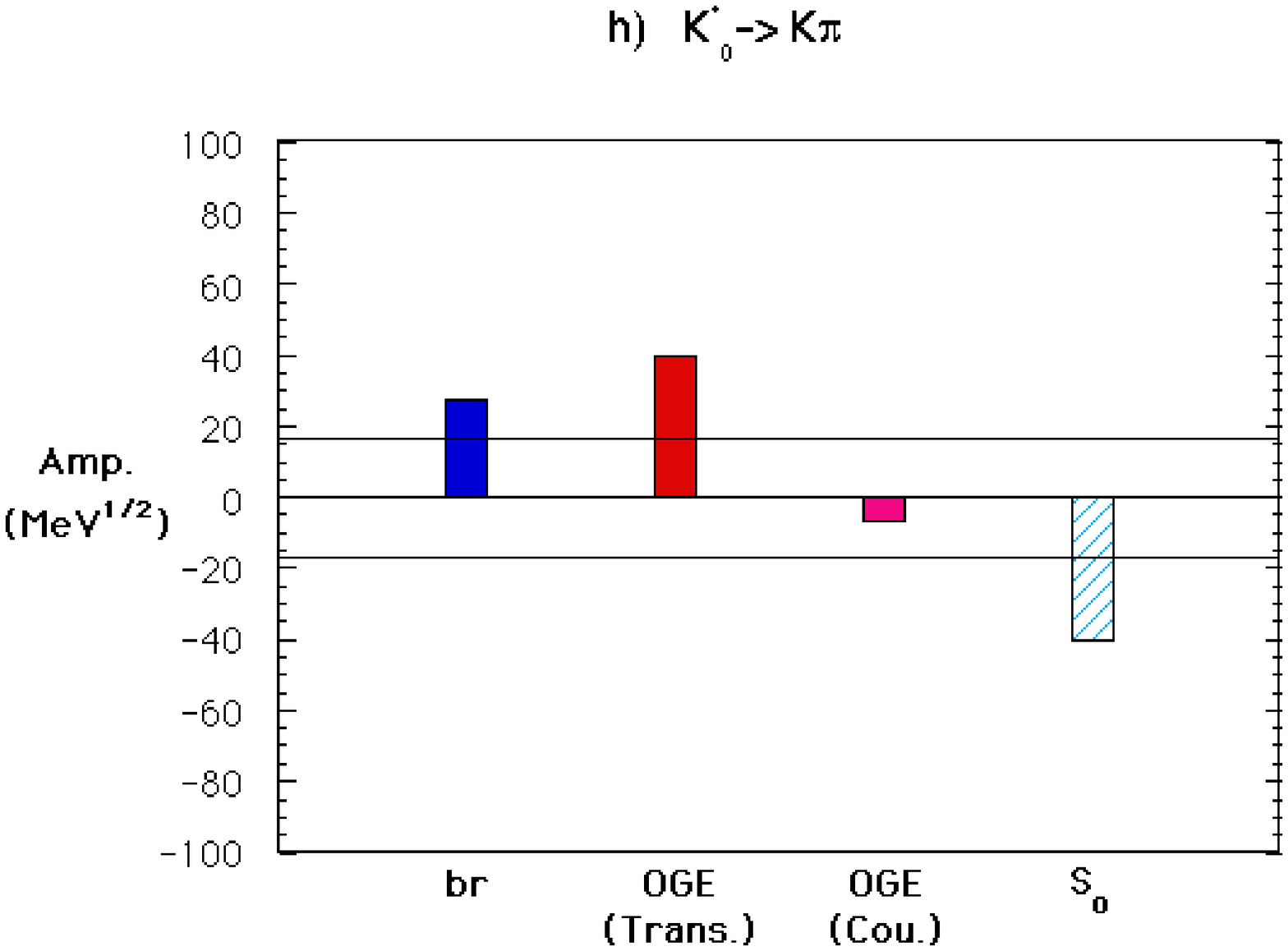}\hss}
\end{figure}
\end{minipage}  
\begin{minipage}[t]{0.5\hsize}
\begin{figure}
\epsfxsize=2.9in
\hbox to \hsize{\hss \epsffile{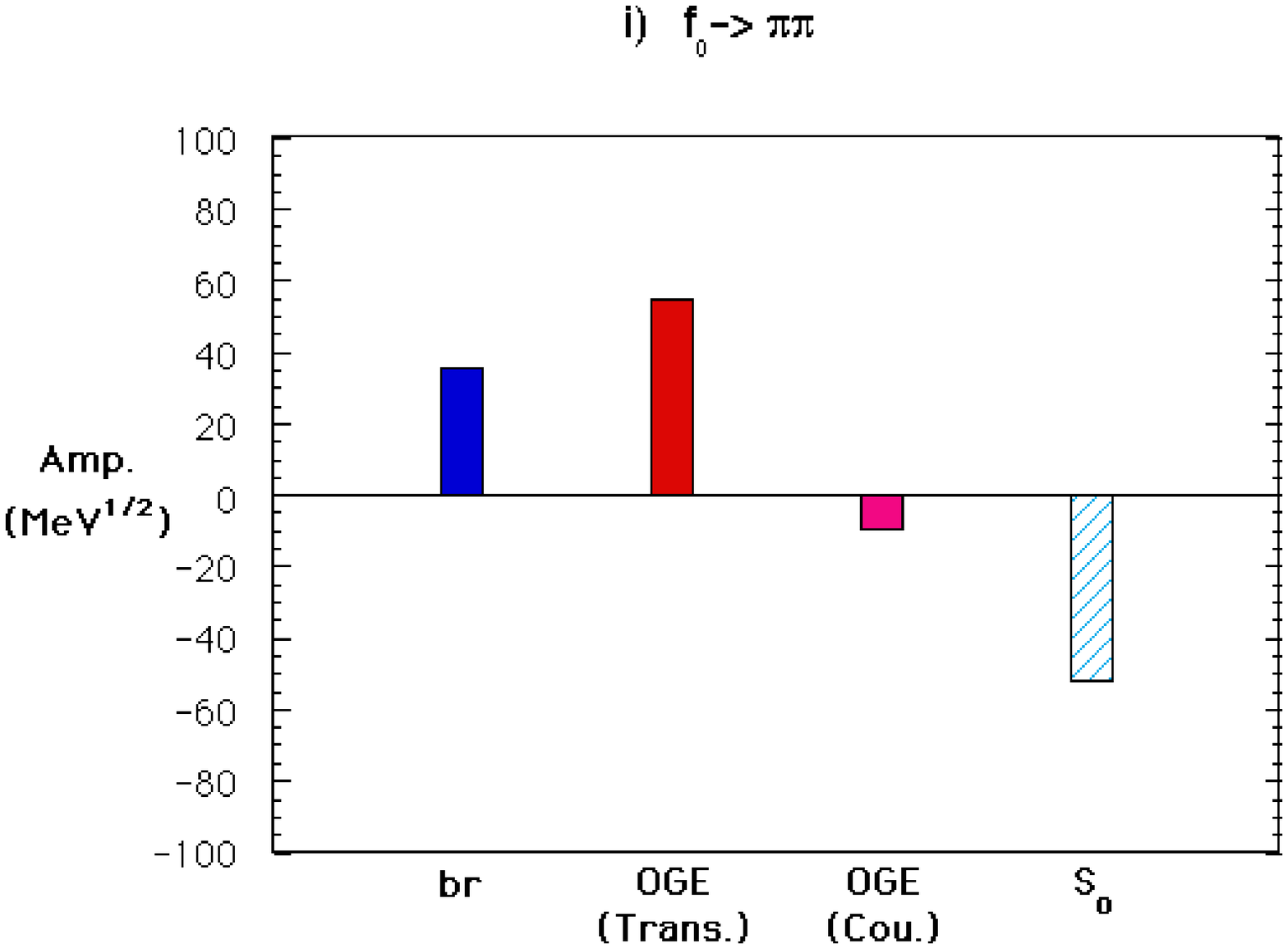}\hss}
\end{figure}
\end{minipage}}

\begin{figure}
{Figure~6a-i.
Decay amplitudes from sKs ($br$ only), 
OGE (transverse) and OGE (coulomb) decay mechanisms,
with parameters $\beta=0.4$~GeV, $b=0.18$ GeV$^2$, $\alpha_s=0.6$ and 
$m_q=0.33$~GeV. The decay amplitude from a constant scalar 
\3p0 term of $S_0=-0.5$~GeV is 
shown
for comparison.
}
\end{figure}

The decay amplitudes due to Coulomb and transverse OGE are also
shown in the histograms.
The Coulomb gluon term is small in all the decays we have considered.
This could have been anticipated; in the limit of a constant kernel $K$ the
Coulomb gluon decay amplitude actually vanishes, because the transition
operator is then proportional to the fermion number operator
$Q=\int d^{\, 3}x j^0$ (squared) and hence cannot pair produce. 
With a slowly varying kernel the Coulomb decay amplitude is nonzero, but
remains small.
Transverse gluon exchange in contrast is
sufficiently large to make
an important contribution in some channels. 
The most notable of these are K$_0^*\to$K$\pi$ and 
$f_0\to\pi\pi$, in which 
the OGE decay amplitude actually dominates the nonperturbative sKs
amplitude
(see figures $6h,i$). 
Recall that the widths of scalar 
$q\bar q$ states are problematical
in the \3p0 model because of a node in their decay amplitudes 
(see figure 2). The large additional
OGE decay 
amplitude insures that the scalar $q\bar q$ states 
will be broad 
resonances, despite the node in the sKs 
or \3p0 decay amplitudes.

Note that the overall 
scale of the total decay amplitudes predicted by the sKs+OGE decay mechanisms 
is too large relative to data by
about a factor of two. 
It is nontrivial that there is even 
approximate agreement; recall in contrast that the overall scale of the
decay rates is not determined in the \3p0 model, but is fitted using the
parameter $\gamma$. 
Since we regard the values
$\alpha_s=0.6$, $b=0.18$~GeV$^2$ and $m_q=0.33$~GeV as reasonably well established
for light quarks, we doubt that
the discrepancy reflects our choices for these parameters. For
this reason we do not show an optimized fit to the experimental rates, which
would give reduced values for $\alpha_s$ or $b$. 

We suggest that the theoretical overestimate of scale may 
indicate the presence of important relativistic corrections
to our nonrelativistic amplitudes,
such as $m/E$ 
factors in (38); if included these would reduce the overall scale of
rates 
and have little effect on the successful
relative rate predictions. 
Other possibile sources of this discrepancy include our choice of
$q\bar q$ wavefunctions (the overall scale of rates 
is very sensitive to the SHO width parameter $\beta$) and
our assumption that confinement may be treated as the exchange of a 
scalar quantum with a linear kernel in configuration space but no
$q$-$q$-$scalar$ form factor.

A large negative constant $S_0$ of up to $\approx -1$~GeV
is often included in potential models, 
and is needed to give a best fit to spectroscopy. 
Of course it may
not actually be present in the scalar potential $S(r)$, and may instead
simply be a way of 
subtracting off a fictitious constituent quark mass contribution 
of $2m_q\approx 0.7$ GeV from the $q\bar q$ spectrum. 
This constant would contribute a decay amplitude 
identical to a \3p0 coupling of
$\gamma = (2^4 / 3^{5/2}) S_0 / m_q $; 
for completeness
we have included such a constant term in our amplitude histograms,
with an intermediate magnitude of $S_0=-0.5$~GeV. 
If an $S_0$ of approximately this magnitude
were actually present
it clearly would make an important contribution, as it is larger than the
$br$ term and opposite in sign.
We find however that any $S_0$ of comparable magnitude leads to 
unrealistic D/S ratios in $b_1\to\omega\pi$ and $a_1\to\rho\pi$.
In both decays D/S passes through zero near $S_0=-0.3$ GeV and diverges
near $S_0=-0.5$ GeV with conventional parameter values 
of $\alpha_s=0.6$, $\beta=0.4$ GeV, $b=0.18$ GeV$^2$ and $m_q=0.33$~GeV, 
whereas $S_0=0$ leads to reasonable D/S values.
Since the D/S ratios are inconsistent with a large negative constant $S_0$ 
we will assume $S_0=0$ subsequently.

\subsection{D/S ratios in JKJ decay models}

In Sec.I.C we noted that the ratio of S- and D-wave amplitudes in the
decays
$b_1\to\omega\pi$
and
$a_1\to\rho\pi$
allowed sensitive tests of the angular quantum numbers of the $q\bar q$
pair produced in a strong decay, and that the ${}^3$P$_0$ model
is in reasonable
agreement with experiment, {\it albeit} for the rather large value of 
$\beta = 0.45$ GeV.
Recall also that
the ratio of
D/S ratios $(a_1/b_1)$ in the $^3$P$_0$ model
is predicted to be $-1/2$ (neglecting minor phase space differences),
independent of the L$_{q\bar q}=0$ and L$_{q\bar q}=1$ radial
wavefunctions.

We find that the sKs interaction leads to very similar predictions to the \3p0 model.
The sKs predictions for D/S
ratios can be read from the decay amplitudes (41-48); for 
$b_1\to\omega\pi$ it is 
\begin{equation}
{a_D\over a_S}\bigg|_{b_1\to\omega\pi}^{\rm sKs} =
-{x^2\over 2^{5/2} }\;
{
\bigg[ \,
{}_1{\rm F}_1 \Big(-{1\over 2}; {3\over 2};\xi\Big) 
+{8\over 15} \, {}_1{\rm F}_1 \Big(-{1\over 2}; {5\over 2};\xi\Big) 
+{8\over 225} \, {}_1{\rm F}_1 \Big(-{1\over 2}; {7\over 2};\xi\Big)
\bigg]
\over
\bigg[  \,
{}_1{\rm F}_1 \Big(-{3\over 2}; -{1\over 2};\xi\Big) 
-{12\over 5}  {}_1{\rm F}_1 \Big(-{3\over 2}; {1\over 2};\xi\Big) 
+ {8\over 15} \, {}_1{\rm F}_1 \Big(-{3\over 2}; {3\over 2};\xi\Big)
\,
\bigg]
}
\end{equation}
This rather complicated result is shown as a 
function of $\beta$ in figure 7,
together with the \3p0 prediction (21) and the corresponding 
$a_1\to \rho\pi$ results. 
The sKs and \3p0 decay models
evidently give remarkably similar D/S ratios, and are essentially
indistinguishable in these reactions.

The ratio of $a_1 / b_1$ D/S ratios with the sKs interaction is 
predicted to be
\begin{equation}
{
{a_D\over a_S}\bigg|_{a_1\to\rho\pi}
\over
{a_D\over a_S}\bigg|_{b_1\to\omega\pi} } = -{1\over 2}
\end{equation}
just as in the \3p0 model.
Thus, to the extent that the \3p0 model is successful in explaining
D/S ratios, this is
a success of the sKs model as well. Some similarity
could have been anticipated since both the \3p0 and sKs decay models
assume pair production
with $^3$P$_0$ quantum numbers, and differ only in the presence
or absence of spatial correlations
with incoming quark lines.

\begin{figure}
$$\epsfxsize=6truein\epsffile{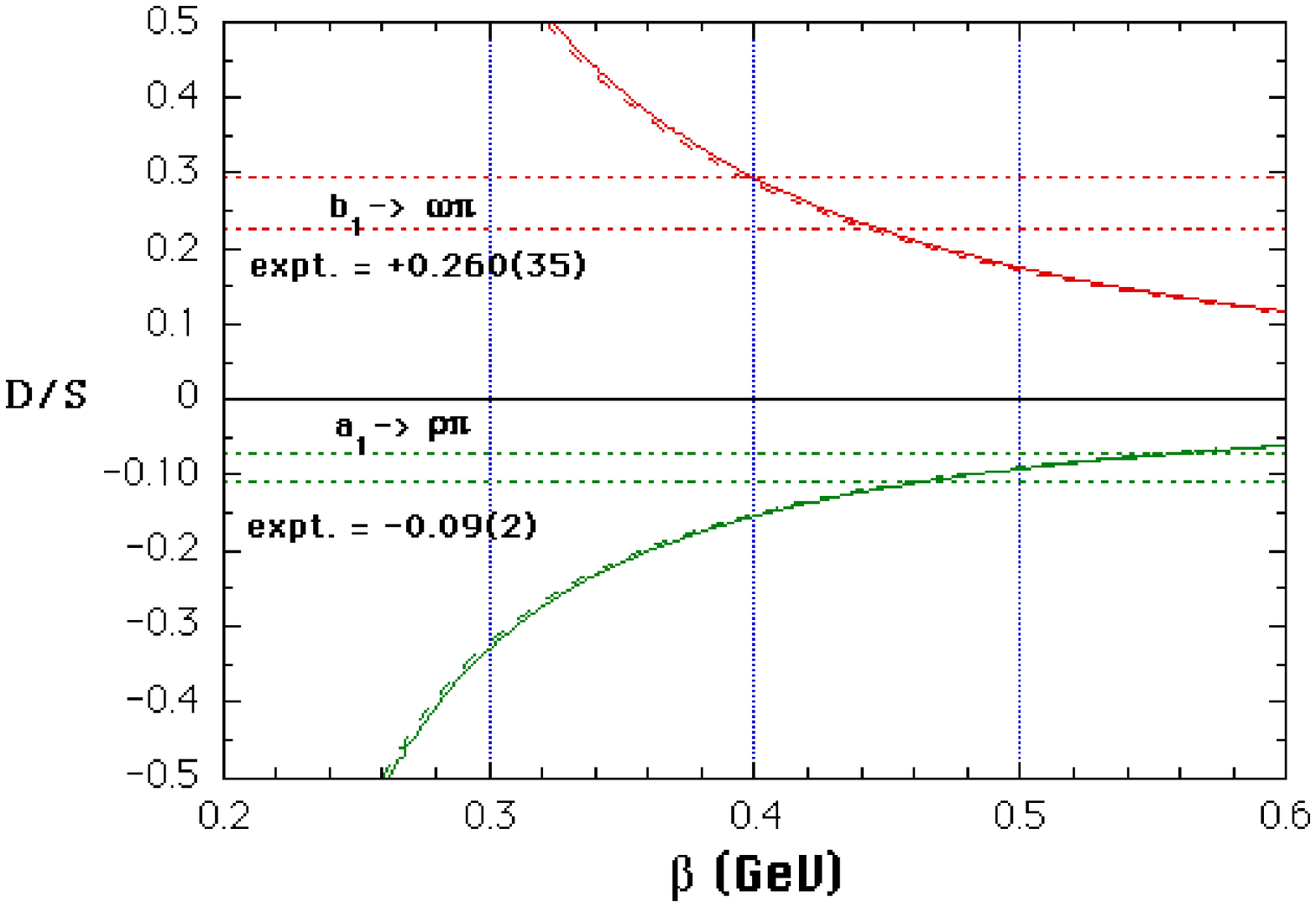}$$
{Figure~7.
D/S ratios in $b_1\to\omega\pi$ and $a_1\to\rho\pi$ 
predicted by the \3p0 model (dashed) and the
sKs model with $S(r)=br$ (solid).
}
\end{figure}

The simple ratio in (53) 
follows in the j$^0$Kj$^0$ model as well, and appears to be
generally true if the initial lines have no spin-flip amplitude
and identical $(a_1,b_1)$ and $(\rho,\omega,\pi)$ 
radial wavefunctions are assumed. In
contrast, the j$^{\rm T}$Kj$^{\rm T}$  transverse OGE
interaction has an initial-line spin-flip
amplitude, so that the ratio departs from $-1/2$. 
This may be useful
as a signature of the OGE component in the
physical decay amplitude.

Although
these D/S ratios are often cited as an argument against an OGE decay mechanism, 
the actual OGE calculation does not appear to have been carried out in the
literature. (We emphasize that OGE is not
equivalent to the ``$^3$S$_1$'' decay model, for which results do exist 
\cite{GS,ABCKP}.) 
We find that the full (Coulomb + transverse) OGE D/S ratios are
given by
\begin{equation}
{a_D\over a_S}\bigg|_{b_1\to\omega\pi}^{\rm OGE} =
-{x^2\over 2^{5/2}3 }\;
{
\bigg[ \,
{}_1{\rm F}_1 \Big({1\over 2}; {3\over 2};\xi\Big) 
+{8\over 9} \, {}_1{\rm F}_1 \Big({1\over 2}; {5\over 2};\xi\Big) 
+{88\over 135} \, {}_1{\rm F}_1 \Big({1\over 2}; {7\over 2};\xi\Big) 
\bigg]
\over
\bigg[  \,
{}_1{\rm F}_1 \Big(-{1\over 2}; -{1\over 2};\xi\Big) 
-{20\over 3} \, {}_1{\rm F}_1 \Big(-{1\over 2}; {1\over 2};\xi\Big) 
+{56\over 9} \,  {}_1{\rm F}_1 \Big(-{1\over 2}; {3\over 2};\xi\Big) 
\,
\bigg]
}
\end{equation}
and
\begin{equation}
{a_D\over a_S}\bigg|_{a_1\to\rho\pi}^{\rm OGE} =
-{x^2\over 2^{9/2}3^2 }\;
{
\bigg[ \,
{}_1{\rm F}_1 \Big({1\over 2}; {5\over 2};\xi\Big) 
+{28\over 45} \, {}_1{\rm F}_1 \Big({1\over 2}; {7\over 2};\xi\Big)
\bigg]
\over
\bigg[  \,
{}_1{\rm F}_1 \Big(-{1\over 2}; {1\over 2};\xi\Big) 
- {10\over 9} \, {}_1{\rm F}_1 \Big(-{1\over 2}; {3\over 2};\xi\Big)
\,
\bigg]
} \ .
\end{equation}
These are shown in figure 8; evidently the hypothesis of OGE dominance of these
decays can indeed be rejected, because the OGE decay mechanism predicts D/S 
amplitude ratios of the wrong sign for both $a_1\to\rho\pi$ and $b_1\to \omega\pi$.
\begin{figure}
$$\epsfxsize=6truein\epsffile{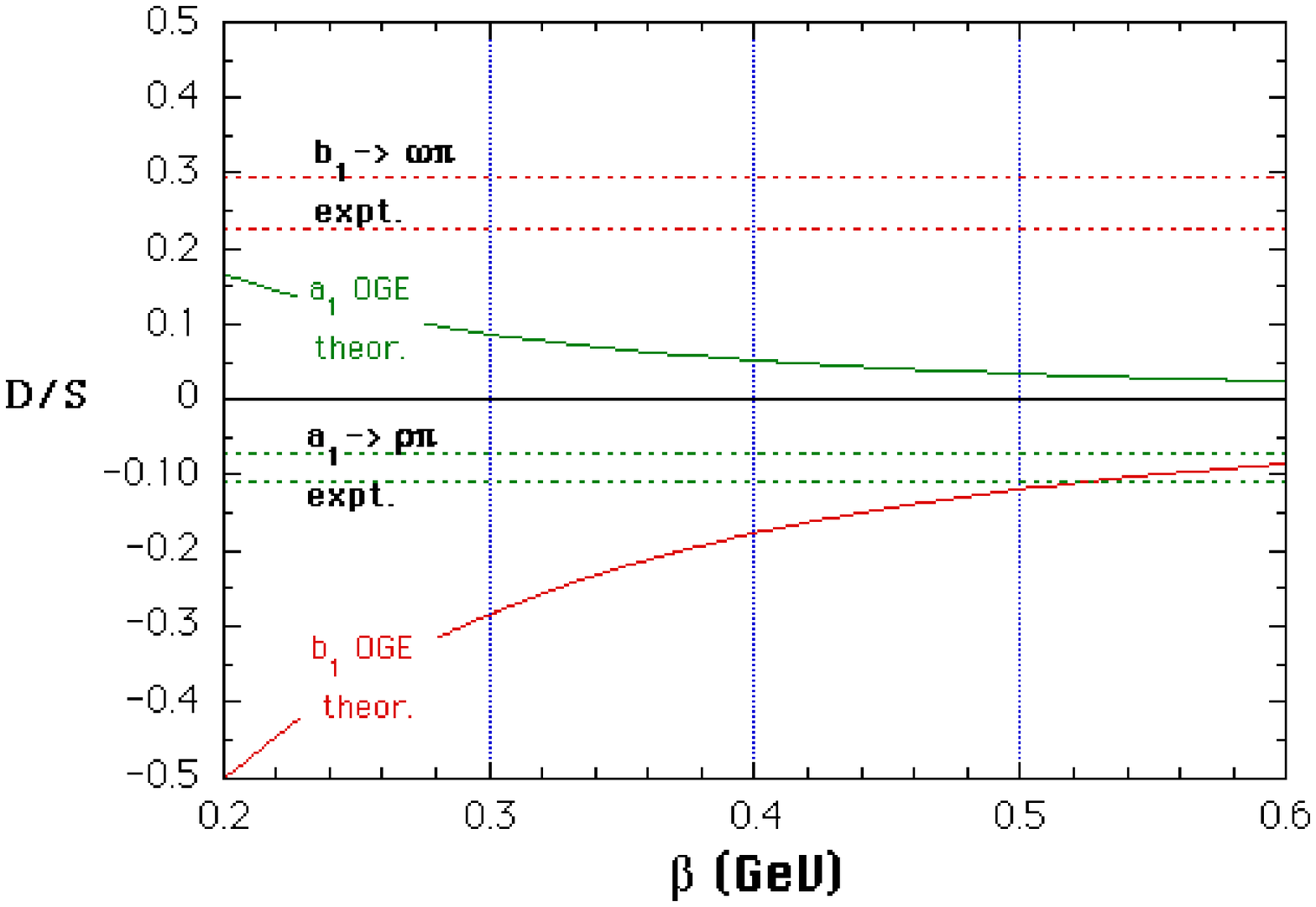}$$
{Figure~8.
D/S ratios in $b_1\to\omega\pi$ and $a_1\to\rho\pi$ assuming pure OGE decay amplitudes.
}
\end{figure}

In reality both sKs and OGE 
decay amplitudes are present, and figure 6 shows
that the OGE contributions are not negligible here. It is especially interesting
to investigate the combined effect of
sKs and OGE amplitudes, because it may be possible to identify the individual
contributions 
through interference. In the following we set
$M(a_1) = M(b_1) = 1.23$~GeV and
$M(\rho) = M(\omega) = 0.78$ GeV 
to remove small phase space effects. First 
in figure~9 we show the D/S ratio in $b_1\to\omega\pi$ as a function of 
the OGE coupling strength $\alpha_s$ for a realistic 
sKs amplitude ($b=0.18$ GeV$^2$), 
for several values of the wavefunction scale $\beta$.
\begin{figure}
$$\epsfxsize=6truein\epsffile{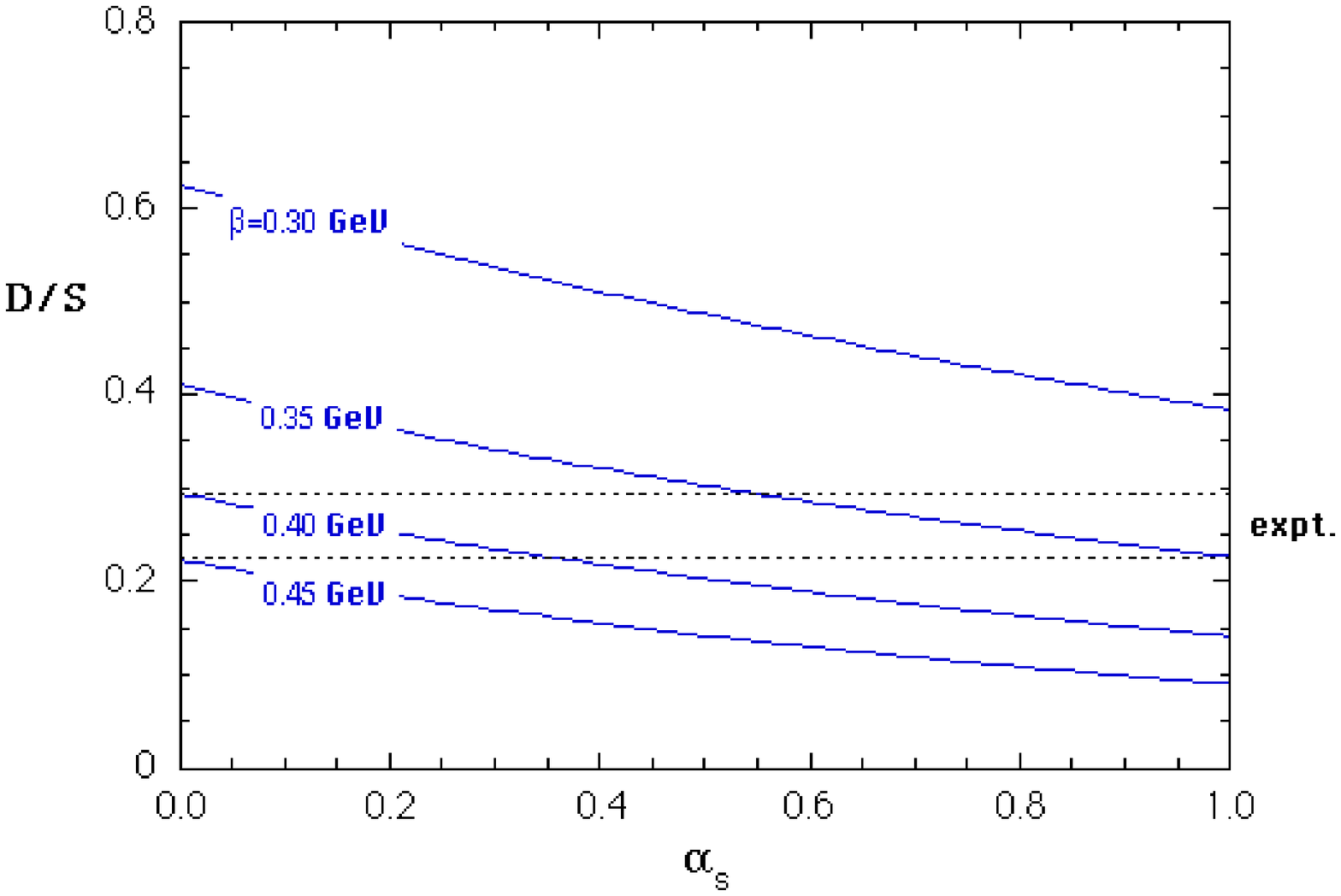}$$
{Figure~9.
D/S ratio in $b_1\to\omega\pi$ with combined sKs and OGE decay amplitudes.
}
\end{figure}
\noindent
Evidently the effect of OGE is to reduce D/S somewhat,
and for the conventional $\alpha_s=0.6$ this leads to a more realistic
value of the wavefunction scale, $\beta \approx 0.35$ GeV.
Recall that a rather large
$\beta\approx 0.45$ GeV was 
required to fit the observed $b_1$ and $a_1$ D/S ratios 
with the sKs interaction treated in isolation. 

A more definitive 
test of the presence of OGE contributions involves the ratio of
$a_1 / b_1$ D/S ratios; 
this is exactly $-1/2$ 
(assuming identical phase space and spatial wavefunctions) 
for any sKs or \3p0 decay strength.
We show this ratio versus the OGE coupling in figure~10, 
for the same parameter set
as figure~9.
\begin{figure}
$$\epsfxsize=6truein\epsffile{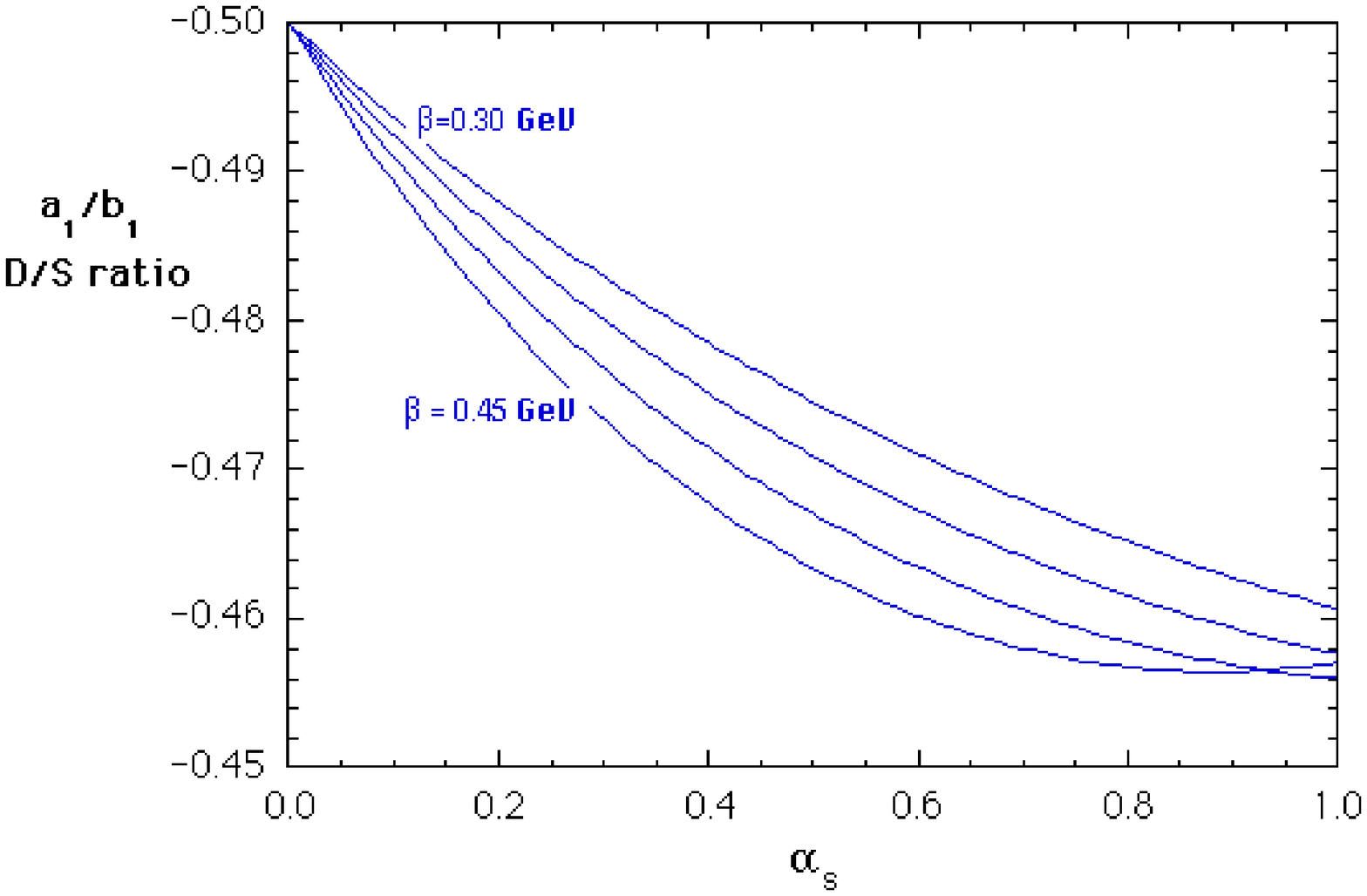}$$
{Figure~10.
The ratio of D/S ratios, $(a_1\to\rho\pi)/(b_1\to\omega\pi)$, 
with combined sKs and OGE decay amplitudes.
}
\end{figure}
\noindent
For $\alpha_s=0.6$ we expect a reduction in the $a_1/b_1$ D/S ratio
to $\approx -0.46$ to $-0.47$. 
The current experimental ratio of $-0.35(9)$ suggests
a larger departure from the sKs and \3p0 prediction of $-1/2$, but 
is consistent with either theoretical result within errors. 
A more accurate experimental
determination of these D/S ratios would be interesting as a test of the expected
OGE corrections to nonperturbative sKs or \3p0 decay amplitudes.

\section{Summary, conclusions and future applications.}

In this paper we have discussed the current theoretical understanding
of open flavor strong decays and proposed new ``microscopic'' quark and 
gluon mechanisms for these decays. We began by reviewing the \3p0 decay
model and introduced a diagrammatic representation for its decay 
amplitudes.
We then carried out
detailed calculations of the decay amplitudes for a representative set of
light meson decays,
first in the \3p0 model and then in microscopic decay models, which assume
that the $q\bar q$ pair
is created by an OGE mechanism or by nonperturbative pair production from
the scalar confining potential. These results were derived analytically
using well established light quark interactions and SHO wavefunctions.

We find that 
with conventional quark model parameters
the light $q\bar q$ open-flavor meson decays are usually dominated
by $q\bar q$ pair production from the scalar confining potential.
We refer to this as the sKs (scalar-kernel-scalar) decay model.
This explains the success of the phenomenological \3p0 decay model, since the
$\bar \psi \psi$ scalar current produces pairs in a \3p0 state, which is assumed
{\it a priori} in the \3p0 model. 
The dimensionless pair production strength $\gamma$ 
is
not determined theoretically in the \3p0 model, 
and is treated as a free parameter. In contrast, 
in our calculations
the absolute decay rates are completely determined by the $q\bar q$ wavefunction
scale $\beta$ 
and known QCD interaction parameters $\alpha_s$, $b$ (string tension)
and $m_q$. In the sKs decay model the \3p0 strength $\gamma$ 
corresponds approximately to the
dimensionless combination $b/m_q\beta$.

Although there are differences in detail between the \3p0 and sKs models
due to the different overlap
integrals, 
we find that the numerical predictions for relative decay rates and 
amplitudes in the decays we have considered 
are remarkably similar.
In addition to explaining the success of the \3p0 model, our description
of decays also
accounts naturally for the absence of ``hairpin'' diagrams; 
since the microscopic pair
production interactions produce $q\bar q$ pairs in color-octet states, hairpin
diagrams are forbidden without additional interactions. In the usual 
$^3$P$_0$-like decay models
one assumes pair production of a $q\bar q$ color singlet, and hairpin diagrams
are simply ignored without theoretical justification. 
Other tests of the color state of $q\bar q$ pair production in decays would be
useful to discriminate between the \3p0 model and our color-octet description.

An unsolved problem is that the overall scale of the amplitudes in the 
full sKs+OGE microscopic decay model
is too large by about a factor of two, given conventional quark model
parameters. This discrepancy of scale
may be due to 
our neglect of relativistic
effects such as the $m_q/E$ external line normalizations in (38).

Although we have shown that OGE $q\bar q$ pair production 
is usually dominated by the nonperturbative sKs decay
amplitudes, 
we do find numerically important
OGE contributions in some
channels, notably $^3$P$_0 \to {}^1$S$_0$+$^1$S$_0$. In that case 
OGE pair production
actually dominates the nonperturbative sKs amplitude, and insures broad
$f_0(1300)$ and K$^*_0(1430)$ $q\bar q$ resonances.
OGE pair production also makes characteristic contributions 
to some observables such
as the D/S amplitude ratios in $b_1\to\omega\pi$ and $a_1\to\rho\pi$. It may
be possible to identify OGE contributions in these and similar multiamplitude
decays.

In future work it would be interesting to apply these microscopic decay calculations
to charmonium (where the transverse OGE contribution should be much smaller) and
to strong decays of light baryons, for which new data is expected from
the CEBAF experimental program. The simple exercise of extending these light meson
calculations to higher-L$_{q\bar q}$ initial mesons
would also be very useful, as it appears likely
that large departures from \3p0 and sKs 
model predictions due to OGE contributions could
be found among the many decay channels available to these states.

\newpage
\acknowledgements

We would like to acknowledge useful communications
with W.Bardeen, H.Blundell, D.V.Bugg, S.Capstick,
F.E.Close, E.Eichten,
P.Geiger, S.Godfrey, N.Isgur, A.LeYaouanc, Z.P.Li, D.Lichtenberg, P.R.Page,
O.Pene, J.M.Richard, W.Roberts, J.Rosner and N.T\"ornqvist. 
We also acknowledge the support of J.Beene
of the ORNL Physics Division, 
and the assistance of M.D.Kovarik with presentation.
This research was sponsored in part by
the United States Department
of Energy under contracts 
DE-FG02-96ER40944 at North Carolina State University and
DE-AC05-840R21400 managed by
Lockheed Martin Energy Systems Inc. at Oak Ridge National Laboratory.

\newpage

\appendix
{}

\section{Diagrammatic formulation of the $^3$P$_0$ model.}

\subsection{General results.}

One may simplify calculations of decay amplitudes in the \3p0 model by
developing a diagrammatic description.
This is nonessential for
$^3$P$_0$ model calculations of
meson decays since there are only
two diagrams, but in our generalized decay models
there are four
diagrams, and it is useful to distinguish their contributions to
prove relations between them. We anticipate that the
diagrammatic description will also be useful in the more
complicated combinatorics of
baryon decays.

We begin by noting that the 
pair production component of the \3p0 
Hamiltonian (2) can be written in terms of creation
operators as
\begin{equation}
H_I =
\sum_{s\bar s} \int \, d^{\, 3} k \; g {m_q\over E_k}
\Big[\, \bar u_{\vec k s} v_{-\vec k \bar s} \Big] \;
b^\dagger_{\vec k s} \,
d^\dagger_{-\vec k \bar s} \ .
\end{equation}
We associate both the coupling constant $g$ and
the external-line spinor bilinear 
${m_q\over E_k} \; [ \bar u_{\vec k s} v_{-\vec k \bar s}] $ with an effective \3p0 
$q\bar q$ pair production vertex.
There is no additional factor of $-i$, unlike conventional field
theoretic Feynman rules, because we are determining the matrix
element of $H_I$ instead of the $T$-matrix.

We assume nonrelativistic $q\bar q$ wavefunctions for the initial and
final mesons; in our notation a meson
state is of the form
\begin{equation}
|A\rangle = \int d^{\, 3}a \int d^{\, 3} \bar a \; \phi(\vec a - \vec {\bar a}\, )
\; \delta( \vec A - \vec a - \vec {\bar a}\, ) \; |a\bar a\rangle \ ,
\end{equation}
with implicit spin
and flavor wavefunctions that are the usual nonrelativistic
quark model forms. (For details of our conventions for the
wavefunctions see reference \cite{BS}, Appendix A.; note that
a factor of $1/(2\pi)^{3/2}$ was inadvertantly omitted
from the normalization of the wavefunction in equation (A15) in that reference.) 

One may now evaluate the Hamiltonian matrix element for
the decay A$\to$BC in terms of straightforward overlap integrals.
Schematically,
we have a matrix element of the type
\begin{equation}
\langle b\bar bc\bar c| \,
b^\dagger_{\vec k s} \,
d^\dagger_{-\vec k \bar s} \,
|a\bar a\rangle \ ,
\end{equation}
and it is useful to distinguish two Feynman diagrams
(figure A1) which we
refer to as $d_1$ if the produced quark goes into meson B and
$d_2$ if it goes into C. 

\begin{figure}
$$\epsfxsize=4truein\epsffile{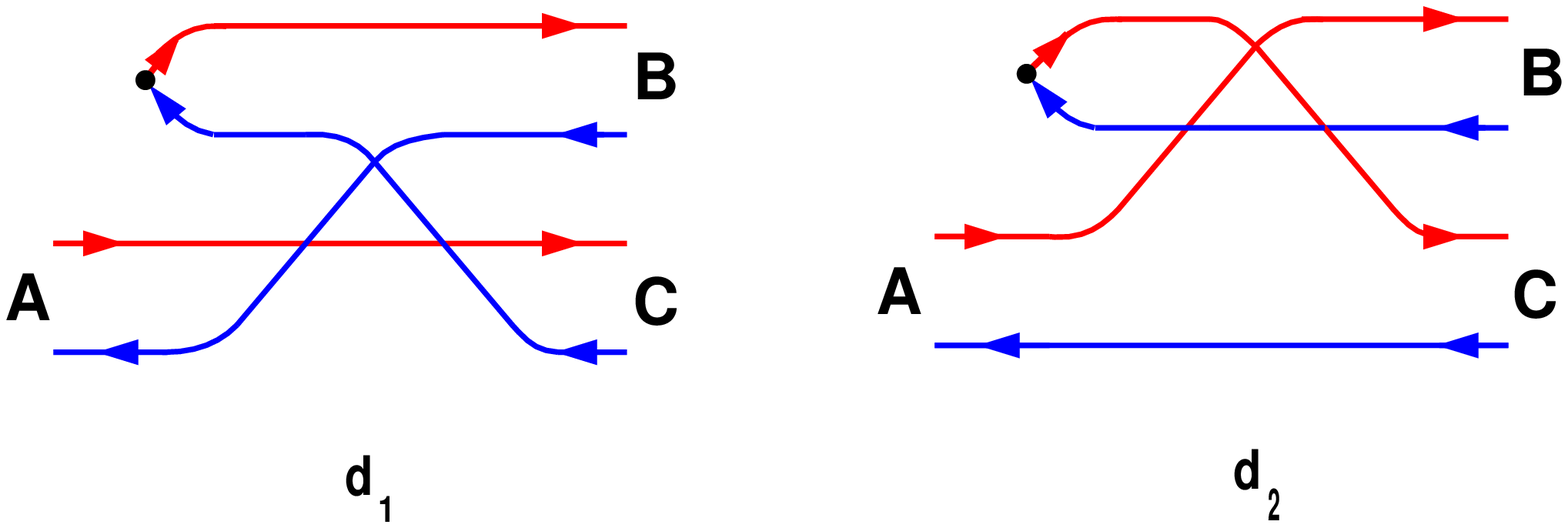}$$
{Figure~A1.
$q\bar q$ meson decay diagrams in the $^3$P$_0$ decay model.
}
\end{figure}

These are drawn so the ordered
quark and antiquark
content of the
state can be read from the diagram by drawing a vertical line through it.
Thus, in $d_1$ the initial state is
$|A\rangle = |a\bar a\rangle$
but immediately after the pair creation we have
$b_q^\dagger d_{\bar q}^\dagger | a \bar a \rangle
 =  | q \bar q a \bar a \rangle $.
This is then rearranged to $|q \bar aa\bar q\rangle$, which is
projected onto the final state $|BC\rangle=|b\bar b c\bar c\rangle$.
The odd number of crossed lines in each diagram
implies a $(-1)$ phase 
(the ``signature" of the matrix element) due to permutation of
quark and antiquark operators.

Specializing to diagram $d_1$, the associated matrix element is 
of the form
\begin{equation}
\langle BC| H_I | A \rangle_{d_1} =
I_{\rm signature} \cdot
I_{\rm flavor} \cdot
{\bf I}_{\rm spin+space} \ .
\end{equation}
The spatial overlap integral associated with
diagram $d_1$ (before the spin matrix element is taken, which gives
${\bf I}_{\rm spin+space}(d_1)$) is 
\begin{displaymath}
{\bf I}_{\rm space}(d_1) =
\int\!
\int\!
\int\!
\int\!
\int\!
\int\!
\, d^{\, 3} a
\, d^{\, 3}\bar a
\, d^{\, 3} b
\, d^{\, 3}\bar b
\, d^{\, 3} c
\, d^{\, 3}\bar c
\end{displaymath}
\begin{displaymath}
\underbrace{
 \phi(\vec a - \vec {\bar a}\, )
\, \delta(\vec A - \vec a - \vec {\bar a}\, )
}_{\rm meson \ A}
\;
\underbrace{
\phi^*(\vec b - \vec {\bar b}\, )
\, \delta(\vec B - \vec b - \vec {\bar b}\, )
}_{\rm meson \ B}
\;
\underbrace{
\phi^*(\vec c - \vec {\bar c}\, )
\, \delta(\vec C - \vec c - \vec {\bar c}\, )
}_{\rm meson \ C}
\end{displaymath}
\begin{equation}
\underbrace{
\int \, d^{\, 3} k \; g{m_q\over E_k}
\Big[\, \bar u_{\vec k s} v_{-\vec k \bar s} \Big]
\, \delta{( \vec k - \vec b \, ) }
\, \delta{( -\vec k - \vec {\bar c} \, ) }
}_{\rm pair\ production\ amplitude}
\
\underbrace{
\, \delta{( \vec a - \vec c \, ) }
\, \delta{( \vec {\bar a} - \vec {\bar b} \, ) }
}_{\rm spectator\  lines}
\end{equation}
This result can be read directly from the diagram (see figure A2), using
the pair production vertex (A1) and the fact that each unscattered
``spectator" line
gives a factor of $\delta(\vec k_i - \vec k_f )$.
\begin{figure}
$$\epsfxsize=6truein\epsffile{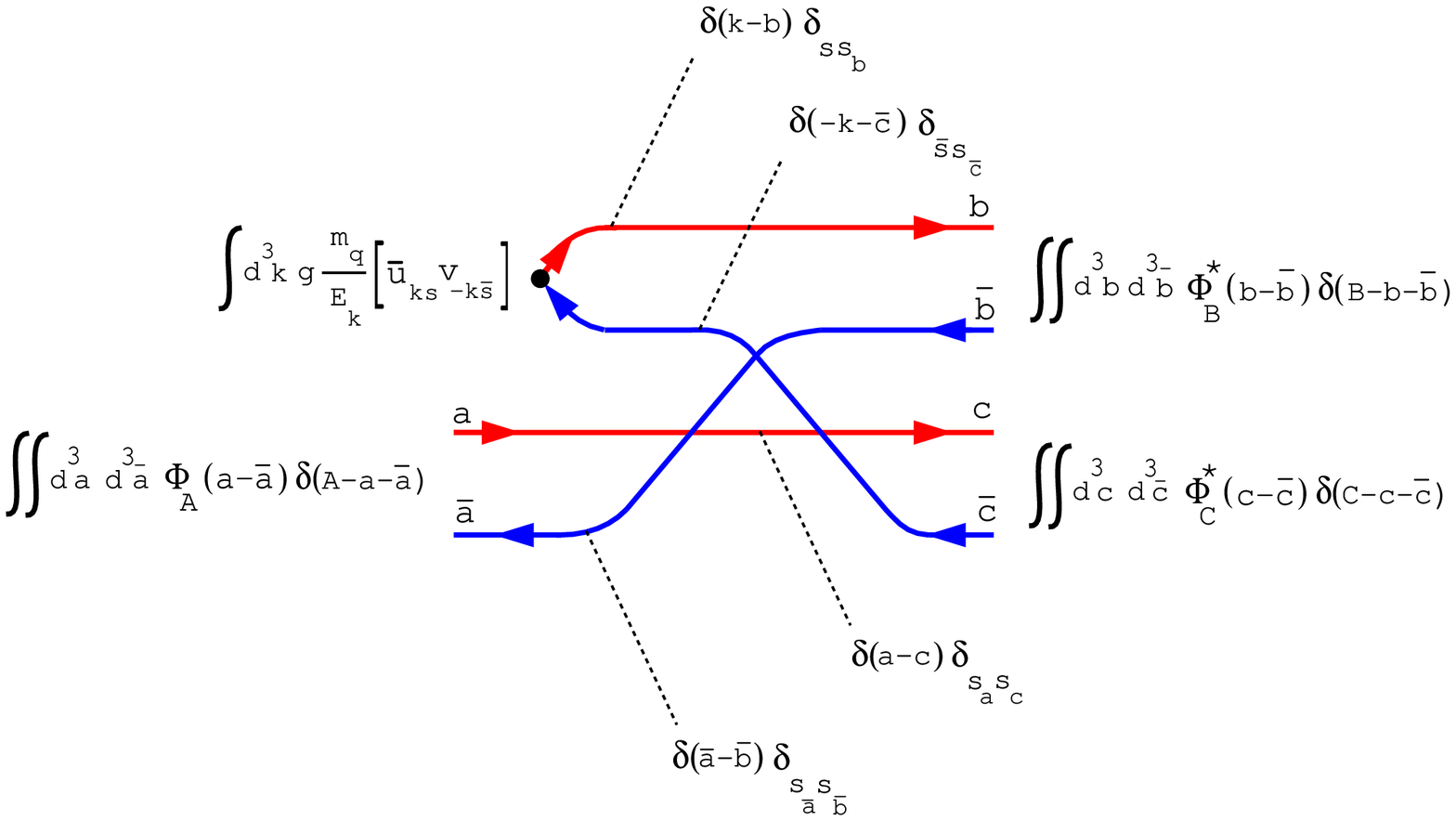}$$
{Figure~A2.
Determination of the spatial overlap integral 
for diagram $d_1$.
}
\end{figure}
\noindent
Thus the decay amplitude is a 21-dimensional integral involving
21 one-dimensional delta functions. Only 18 of the integrations can
be carried out trivially, which leaves a three-dimensional overlap
integral times a $\delta(\vec A - \vec B - \vec C\, )$
momentum-conserving delta function, 
\begin{equation}
{\bf I}_{space} = I_{space} \; \delta(\vec A - \vec B - \vec C \; ) \ ,
\end{equation}
as in (4). These overlap
integrals
are explicitly (setting $\vec A=0$ and $\vec B = -\vec C$)

\begin{equation}
I_{space} (d_1) = \int \, d^{\, 3} k \;
\phi_A(2\vec k - 2\vec B ) \,
\phi_B^*(2\vec k - \vec B ) \,
\phi_C^*(2\vec k - \vec B ) \,
\cdot g {m_q\over E_k}
\Big[ {\bar u}_{\vec k s_b} v_{-\vec k s_{\bar c}} \Big]
\ ,
\end{equation}
\begin{equation}
I_{space} (d_2) = \int \, d^{\, 3} k \;
\phi_A(2\vec k + 2 \vec B) \,
\phi_B^*(2\vec k + \vec B ) \,
\phi_C^*(2\vec k + \vec B ) \,
\cdot g {m_q\over E_k}
\Big[ {\bar u}_{\vec k s_c} v_{-\vec k s_{\bar b}} \Big] \ .
\end{equation}
The spin factor
and labels $s_q,s_{\bar q}$ in these overlaps depend on the reaction
being considered, and are determined by
the external line labels
attached to the diagrams, as illustrated below.

\subsection{An illustrative $^3$P$_0$ decay: $\rho\to\pi\pi$.}

Here and in Appendix C we will use the decay
$\rho^+(+\hat z)\to \pi^+\pi^o$ to illustrate our techniques, and 
simply quote results for other cases in the text. 
For this decay the flavor states
are
$|\rho^+\rangle =
|\pi^+\rangle = -|u\bar d\, \rangle$ and
$|\pi^o\rangle = (|u\bar u\, \rangle - |d\bar d\, \rangle )/\sqrt{2}$. 
Figure A3 shows the evaluation of the flavor factor for diagram $d_1$.

\begin{figure}
$$\epsfxsize=4truein\epsffile{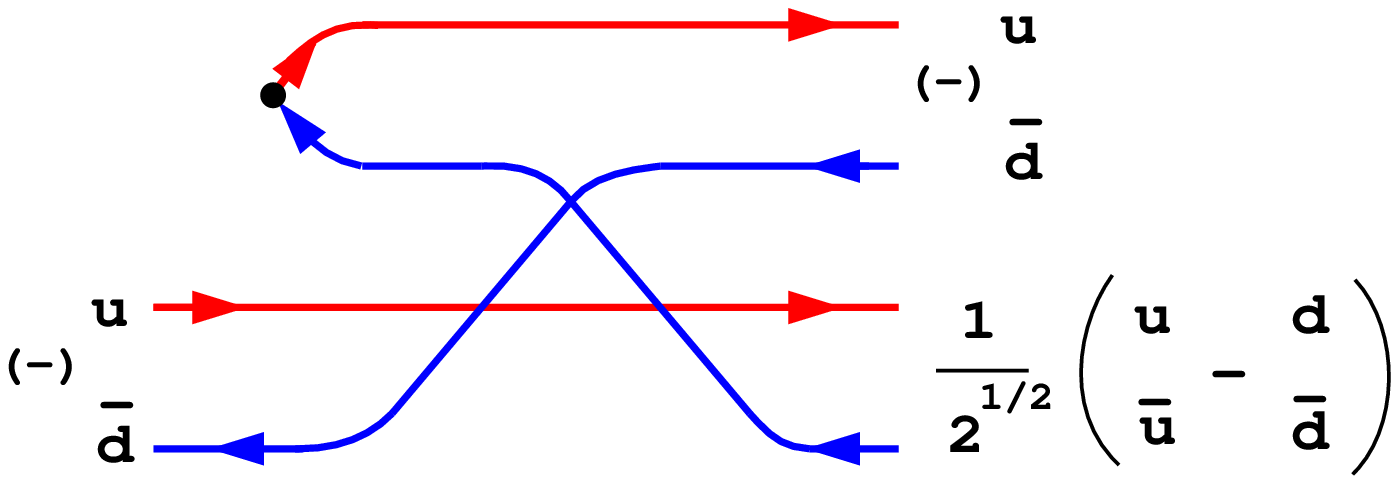}$$
{Figure~A3.
Determination of the flavor factor of $+1/\sqrt{2}$ 
for diagram $d_1$ in $\rho^+\to\pi^+\pi^o$.
}
\end{figure}

The flavor factors associated with the other decays we consider 
can be determined similarly,
and are tabulated below. For a given decay, such as $f_2\to\pi\pi$, we give the
flavor weight factors for each diagram in a specific charge channel 
(in this case $f_2\to\pi^+\pi^-$). The resulting decay rate for
$f_2\to\pi^+\pi^-$ must then be multiplied by a multiplicity factor 
${\cal F}$ to sum over all
final flavor states (here ${\cal F}=3/2$ for $\pi^+\pi^-$ and $\pi^o\pi^o$). 
\vskip 1cm

\begin{center}
\begin{tabular}{||c|c|c|c|c||}  \hline
\multicolumn{5}{||c||}{Table A1. Flavor Weight Factors.} \\ \hline 
$\ $ Generic Decay $\ $  & $\ $ Subprocess $\ $ & $\ I_{flavor}(d_1)\ $ 
& $\ I_{flavor}(d_2)\ $  & 
$\ {\cal F}\ $   
\\ \hline
$\rho \to \pi\pi$ & $\rho^+ \to \pi^+\pi^o$ & $+1/\sqrt{2}$ & $-1/\sqrt{2}$    &  $1$
\\  \hline
$f \to \pi\pi$ & $f \to \pi^+\pi^-$ & $-1/\sqrt{2}$ & $-1/\sqrt{2}$    &  $\  3/2\  $ 
\\  
\hline
$a \to \rho\pi$ & $a^+ \to \rho^+\pi^o$ & $+1/\sqrt{2}$ & $-1/\sqrt{2}$    &  $2$
\\  \hline
$b \to \omega\pi$ & $b^+ \to \omega\pi^+$ & $+1/\sqrt{2}$ & $+1/\sqrt{2}$    &  $1$
\\  \hline
$h \to \rho\pi$ & $h \to \rho^+\pi^-$ & $-1/\sqrt{2}$ & $-1/\sqrt{2}$    &  $3$
\\  \hline
${\rm K}^*\to {\rm K}\pi$ & ${\rm K}^{*+}\to {\rm K}^+\pi^o$ 
& $+1/\sqrt{2}$ & $0$  &  $3$  \\  \hline
\end{tabular}
\end{center}

\vskip 1cm
\noindent
The ``flavorless'' decay amplitudes quoted in the text, equations (7-15) and
(40-48), correspond
to
unit flavor factors, 
$I_{flavor}(d_1)=+1$ and $I_{flavor}(d_2)=\pm 1$. (The sign is
chosen so the contributions of $d_1$ and $d_2$ 
add rather than cancel.) To convert these to physical
decay amplitudes one should multiply (7) or (40) by $I_{flavor}(d_1)$ from Table A1,
and multiply the resulting total decay rate by ${\cal F}$. 
The K$^*$ is a special case because only one diagram contributes; in this case
(7) or (40) should instead be multiplied by $I_{flavor}(d_1)/2$.

The
spin states are $|\rho(+\hat z)\rangle = |\uparrow\bar\uparrow\; \rangle$ and
$|\pi\rangle = (|\uparrow\bar\downarrow\; \rangle -
|\downarrow\bar\uparrow\; \rangle)/\sqrt{2}$;
taking the spin matrix element analogously to figure A3, we find a spin
factor of 
$[(-1/2) \bar u_{\vec k, \downarrow }
v_{-\vec k, \downarrow }]$ for this diagram. Combining these results and
using the explicit Dirac spinor matrix element in Appendix B, we find

\begin{equation}
h_{fi}(d_1) =
\underbrace{I_{\rm signature}}_{(-1)} \cdot
\underbrace{I_{\rm flavor}}_{(+{1/ \sqrt{2}} )} \cdot
I_{\rm spin+space}(d_1) 
\end{equation}
where
\begin{equation}
I_{spin+space} (d_1) =  -{1\over 2}g \int d^{\, 3}k \,
\phi_A(2\vec k - 2 \vec B \, )
\;
\phi_B^*(2\vec k -  \vec B \, )
\;
\phi_C^*(2\vec k -  \vec B \, ) \, 
{ (k_x + ik_y) \over E_k} \ .
\end{equation}

The second diagram $d_2$ has an opposite flavor factor, so the
full result is
\begin{equation}
h_{fi} = -{1\over \sqrt{2}} \bigg(
I_{spin+space}(d_1) -
I_{spin+space}(d_2) \bigg) \ .
\end{equation}
Since the overlap integrals satisfy
$I(d_2, -\vec B) = + I(d_1, \vec B )$, $h_{fi}$ is odd under
$\Omega_B \to - \Omega_B$, and can be written as
\begin{displaymath}
h_{fi}
= -\sqrt{2} \, I_{spin+space} (d_1) \hskip 5cm
\end{displaymath}
\begin{equation}
=  +{1\over \sqrt{2} }\; {g\over m_q} \int \, d^{\, 3}k \,
\phi_A(2\vec k - 2 \vec B \, ) \;
\phi_B^*(2\vec k -  \vec B \, ) \;
\phi_C^*(2\vec k -  \vec B \, )\;  ( k_x + ik_y ) \ .
\end{equation}
We have also substituted
$m_q$ for $E_k$ in (A10)
to recover
the nonrelativistic limit, which gives the
$^3$P$_0$ model.

Frequently in the literature these
decay amplitudes are evaluated using SHO
$q\bar q$ wavefunctions, which leads to closed-form results.
For L$_{q\bar q}=0$ mesons our Gaussian momentum space wavefunction
$\phi(\vec k_q - \vec k_{\bar q} )$, with 
$\vec k = \vec k_q = -\vec k_{\bar q}$, 
is
\begin{equation}
\phi ( 2\vec k \, ) = {1\over \pi^{3/4} \beta^{3/2}}\;  e^{-k^2 / 2 \beta^2}
\ .
\end{equation}
We note in passing that the L$_{q\bar q}=1$ wavefunction used in this paper is
\begin{equation}
\phi_{1m} ( 2\vec k \, ) 
=  {2^{3/2}\over 3^{1/2}} \, {1\over \pi^{1/4} \beta^{3/2}}\; {k\over \beta} 
\; e^{-k^2 / 2 \beta^2} \; Y_{1m}(\Omega_k)
\ .
\end{equation}
On substitution into (A12), $\phi(2\vec k \, )$
gives a $\rho^+(+\hat z)\to \pi^+\pi^o$ decay amplitude of
\begin{equation}
h_{fi} = - {2^{7/2} \over 3^3}\;  \pi^{-1/4}
\, {g \over m_q }
\; {P \over
\beta^{3/2}} \;
e^{-P^2/ 12 \beta^2} \, 
Y_{11}(\Omega_B) \, 
\end{equation}
and an ${\cal M}^{\rho^+\to\pi^+\pi^o}_{10}$ amplitude of
\begin{equation}
{\cal M}^{\rho^+\to\pi^+\pi^o}_{10} = - {2^{9/2} \over 3^3}\;  
{1\over \pi^{1/4}\beta^{1/2} } \; 
{g \over 2 m_q }\;
{P \over
\beta} \;
e^{-P^2/ 12 \beta^2} \, 
Y_{11}(\Omega_B) \, 
\end{equation}
Substitution into (6) then gives the total decay rate
\begin{equation}
\Gamma_{\rho\to\pi\pi} =  \pi^{1/2}\; \bigg({2^{10} \over 3^6 }\bigg)
\; 
\bigg({g\over 2m_q}\bigg)^2 \; 
M_\rho \;
\Big( P/ \beta \Big)^3 \,
e^{-P^2/6\beta^2 }
\end{equation}
which with the identification $\gamma = g/2m_q$ 
is the usual \3p0 result.

\vskip .5cm
Many features of this decay rate could be anticipated
on general grounds. Specifically:
\begin{enumerate}
\item  The $(g/m_q)^2$ dependence follows from the $O(g)$ of the amplitude
and the nonrelativistic limit
of (2).
\item   The factor of $M_\rho $ comes from phase space
and the relation $E_B E_C/M_A = M_\rho /4$.
\item  The $P^3$ threshold dependence is expected
for a P-wave final state.
\item   Dimensional arguments then require a factor of $1/\beta^3$.
\item  An exponential in $(P/\beta)^2$ is expected given
SHO wavefunctions.
\end{enumerate}
Only the overall numerical coefficient and the
factor of $1/6$ in the exponential (present in all \3p0
SHO A$\to$BC $q\bar q$ meson decays)
require detailed calculation.

\newpage

\section{Overlap integrals and spin factors.}

In the text we discussed the evaluation of only one of the four
decay diagrams present in A$\to$BC meson decay in JKJ models.
Here we give the overlap integrals associated with all four diagrams:

\begin{displaymath}
I_{space} (d_{1q}) =
+{1\over (2\pi )^3 }
\int \!  \int \;
d^{\, 3}a\,  d^{\, 3}c \;
\phi_A(2\vec a) \;
\phi_B^*(2\vec a + \vec B ) \;
\phi_C^*(2\vec c + \vec B )
\end{displaymath}
\begin{equation}
\Big[ \bar u_{b s_b} \Gamma v_{{\bar c} s_{\bar c}} \Big]
\; {\cal K}(\vec a - \vec c \, ) \;
\Big[ \bar u_{c s_c} \Gamma u_{a s_a} \Big]
\end{equation}
(with implicit momentum constraints specific to this diagram of
$\vec b =  \vec a + \vec B $ and
$\vec{\bar c} = -\vec B - \vec c$),
\begin{displaymath}
I_{space} (d_{1\bar q}) =
-{1\over (2\pi )^3 }
\int \!  \int \;
d^{\, 3}a\,  d^{\, 3}b \;
\phi_A(2\vec a  ) \;
\phi_B^*(2\vec b -  \vec B ) \;
\phi_C^*(2\vec a + \vec B )
\end{displaymath}
\begin{equation}
\Big[ \bar u_{b s_b} \Gamma v_{{\bar c} s_{\bar c}} \Big]
\; {\cal K}(\vec{ \bar a} - \vec {\bar b}  \, )
\;
\Big[ \bar v_{\bar a s_{\bar a}} \Gamma v_{\bar b s_{\bar b}} \Big]
\end{equation}
(with $\vec {\bar c} = -\vec B -\vec c$,
$\vec {\bar a} = -\vec a$ and
$\vec {\bar b} = \vec B - \vec b$),
\begin{displaymath}
I_{space} (d_{2q}) =
+{1\over (2\pi )^3 }
\int \!  \int \;
d^{\, 3}a\,  d^{\, 3}b \;
\phi_A(2\vec a  ) \;
\phi_B^*(2\vec b - \vec B ) \;
\phi_C^*(2\vec a - \vec B )
\end{displaymath}
\begin{equation}
\Big[ \bar u_{c s_c} \Gamma v_{{\bar b} s_{\bar b}} \Big]
\; {\cal K}(\vec a - \vec b \, ) \;
\Big[ \bar u_{b s_b} \Gamma u_{a s_a} \Big]
\end{equation}
(with $\vec  c = \vec a -\vec B$ and
$\vec {\bar b} = \vec B - \vec b$),
\begin{displaymath}
I_{space} (d_{2\bar q}) =
-{1\over (2\pi )^3 }
\int \!  \int \;
d^{\, 3}a\,  d^{\, 3}c \;
\phi_A(2\vec a  ) \;
\phi_B^*(2\vec a  - \vec B ) \;
\phi_C^*(2\vec c + \vec B )
\end{displaymath}
\begin{equation}
\Big[ \bar u_{c s_c} \Gamma v_{{\bar b} s_{\bar b}} \Big]
\; {\cal K}(\vec {\bar a} - \vec {\bar c} \, ) \;
\Big[ \bar v_{\bar a s_{\bar a}} \Gamma v_{\bar c s_{\bar c}} \Big]
\end{equation}
(with $\vec {\bar a} =  - \vec a$,
$\vec {\bar b} = \vec B - \vec a$ and
$\vec {\bar c} = -\vec B - \vec c$\,).

In our evaluation of decay matrix elements in JKJ models we
also require spin matrix elements, which involve
the nonrelativistic $O(p/m)$ matrix elements of Dirac bilinears with
$\Gamma = \gamma^0, \vec \gamma$ and I and Pauli spin matrix elements
for various spin states. These are

\begin{equation}
\lim_{v/c\to 0} \
\Big[ \bar u_{q' s'} \Gamma u_{q s} \Big] =
\left\{
\begin{array}{ll}
\delta_{ss'}
\ &   \Gamma = \gamma^0 \\
{1\over 2m_q} \Big[
(\vec q + {\vec q}\, ' \, ) \delta_{ss'}
-i\langle s'|\vec \sigma \, | s \rangle \,
\times (\vec q\, ' - \vec q \, ) \Big] 
\ &   \Gamma = \vec \gamma \\
\delta_{ss'}
\ &   \Gamma = {\rm I} \\
\end{array}
\right.
\end{equation}

\begin{equation}
\lim_{v/c\to 0} \
\Big[ \bar u_{q s} \Gamma v_{\bar q \bar s} \Big] =
\left\{
\begin{array}{ll}
{1\over 2m_q} \langle s\bar s |\vec \sigma | 0 \rangle 
\cdot (\vec {\bar q} + \vec  q)
\ &   \Gamma = \gamma^0 \\
\langle s\bar s |\vec \sigma \, | 0 \rangle
\ &   \Gamma = \vec \gamma \\
{1\over 2m_q} \langle s\bar s |\vec \sigma | 0 \rangle 
\cdot (\vec {\bar q} - \vec  q)
\ &   \Gamma = {\rm I} \\
\end{array}
\right.
\end{equation}

\begin{equation}
\lim_{v/c\to 0} \
\Big[ \bar v_{\bar q \bar s} \Gamma v_{{\bar q}' {\bar s}'} \Big] =
\left\{
\begin{array}{ll}
\delta_{{\bar s}{\bar s}'}
\ &   \Gamma = \gamma^0 \\
{1\over 2m_q} \Big[
(\vec {\bar q} + {\vec {\bar q}}\, ' \, ) \delta_{{\bar s} {\bar s}'}
-i\langle {\bar s}'|\vec \sigma \, | {\bar s} \rangle \,
\times (\vec {\bar q}\, ' - \vec {\bar q} \, ) \Big] 
\ &   \Gamma = \vec \gamma \\
-\delta_{{\bar s}{\bar s}'}
\ &   \Gamma = {\rm I} \\
\end{array}
\right.
\end{equation}

The explicit scattering and pair-production
matrix elements of Pauli spinors
in terms of
the spherical basis vectors
$\hat e_{\pm} = \mp (\hat x + i\hat y)/\sqrt{2}$ and $\hat e_0=\hat z$,
are
\begin{equation}
\langle \, \uparrow \, |\; \vec \sigma \; |\, \uparrow \, \rangle =
- \langle \, \downarrow \, |\; \vec \sigma \;
|\, \downarrow \, \rangle = \hat e_0
\end{equation}

\begin{equation}
\langle \, \uparrow \, |\; \vec \sigma \; |\, \downarrow \, \rangle =
\sqrt{2} \; \hat e_-
\end{equation}

\begin{equation}
\langle \, \downarrow \, |\; \vec \sigma \; |\, \uparrow \, \rangle =
-\sqrt{2} \; \hat e_+
\end{equation}

\begin{equation}
\langle \, \bar \uparrow \, |\; \vec \sigma \; |\, \bar \uparrow \, \rangle =
- \langle \, \bar \downarrow \, |\; \vec \sigma \;
|\, \bar \downarrow \, \rangle = -\hat e_0
\end{equation}

\begin{equation}
\langle \, \bar \uparrow \, |\; \vec \sigma \; |\, \bar \downarrow \, \rangle =
\sqrt{2} \; \hat e_+
\end{equation}

\begin{equation}
\langle \, \bar \downarrow \, |\; \vec \sigma \; |\, \bar \uparrow \, \rangle =
-\sqrt{2} \; \hat e_-
\end{equation}

\begin{equation}
\langle \, \uparrow\bar\uparrow \, |\; \vec \sigma \; |\, 0 \,  \rangle =
\sqrt{2} \; \hat e_-
\end{equation}

\begin{equation}
\langle \, \uparrow\bar\downarrow \, |\; \vec \sigma \; |\, 0 \,  \rangle =
\langle \, \downarrow\bar\uparrow \, |\; \vec \sigma \; |\, 0 \,  \rangle =
- \hat e_0
\end{equation}

\begin{equation}
\langle \, \downarrow\bar\downarrow \, |\; \vec \sigma \; |\, 0 \, \rangle =
\sqrt{2} \; \hat e_+
\end{equation}

\newpage

\section{$\rho\to\pi\pi$ in JKJ decay models}

We will illustrate a JKJ decay calculation in detail
using the reaction $\rho^+(+\hat z)\to\pi^+\pi^o$,
as in our discussion of the \3p0 model in Appendix A.
We will evaluate the $\rho\to\pi\pi$ decay rate
using the j$^0$Kj$^0$ interaction in (27-29), and simply quote results
in the text for the other cases. 
For each type of JKJ interaction we find that the four diagrams 
give equal contributions,
so we discuss only diagram $d_{1q}$, shown below, in detail.
\begin{figure}
$$\epsfxsize=3truein\epsffile{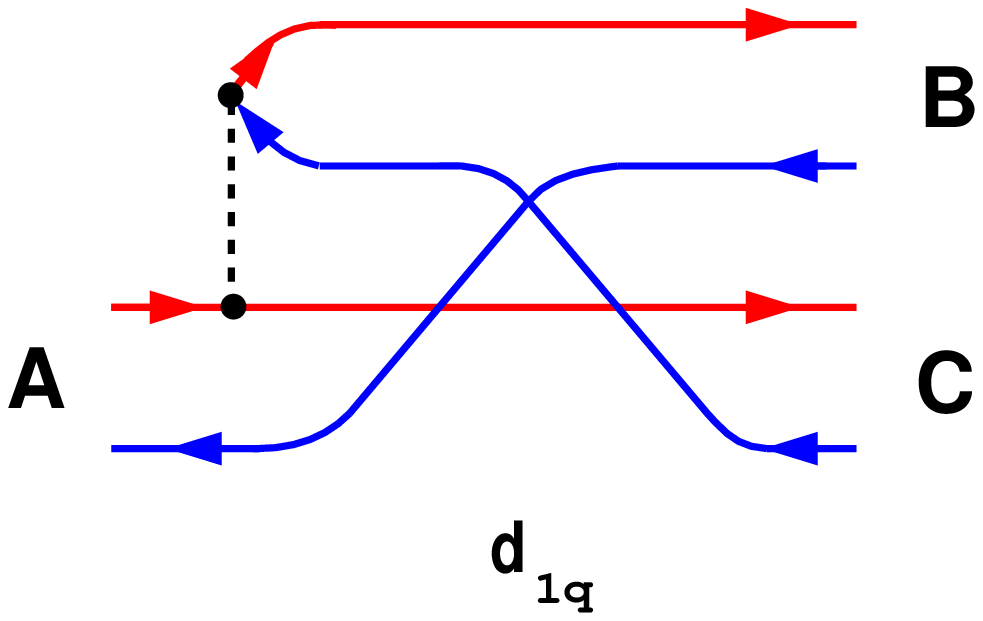}$$
{Figure~C1.
Decay diagram $d_{1q}$.
}
\end{figure}

For all the JKJ interactions we consider this diagram
has a signature of $(-1)$, a color factor of $+2^2/ 3^{3/2}$, and
a flavor factor (for $\rho^+\to\pi^+\pi^o$) of
$+1/2^{1/2}$. Differences in the relative importance of the
decay mechanisms
arise in the spin-space overlap integrals and the
$\Gamma$ spinor matrix elements.

In the text we derived the spatial overlap integral for diagram 
$d_{1q}$ in the nonrelativistic limit, equation (39);
\begin{displaymath}
I_{space} (d_{1q}) =
{1\over (2\pi )^3 }
\int \!  \int \;
d^{\, 3}a\,  d^{\, 3}c \;
\phi_A(2\vec a) \;
\phi_B^*(2\vec a + \vec B ) \;
\phi_C^*(2\vec c + \vec B )
\end{displaymath}
\begin{equation}
\Big[ \bar u_{b s_b} \Gamma v_{{\bar c} s_{\bar c}} \Big]
\; {\cal K}(|\vec a - \vec c \, |) \;
\Big[ \bar u_{c s_c} \Gamma u_{a s_a} \Big] \ .
\end{equation}
To evaluate this for $\rho^+(+\hat z) \to \pi^+\pi^o$, note that 
for the color Coulomb interaction $\Gamma=\gamma^0$ the
spinor bilinears give
\begin{equation}
\Big[ \bar u_{b s_b} \gamma^0 v_{{\bar c} s_{\bar c}} \Big]
\;
\Big[ \bar u_{c s_c} \gamma^0 u_{a s_a} \Big]
= \, {1\over 2 m_q} \;
\langle \, s_b  s_{\bar c} \, | \vec \sigma \, | \, 0 \, \rangle\,
\cdot (\vec b + \vec {\bar c } \, )
\ \delta_{s_c s_a}
\end{equation}
Attaching the spin wavefunctions to the diagram $d_{1q}$ gives an overall
factor of $(-1/2)$ and sets $s_b=\downarrow$ and
$s_{\bar c}=\bar \downarrow$, so the spin matrix element is
$\langle \vec \sigma \rangle_{q\bar q} = 
(-1/2) \langle  \downarrow \bar \downarrow  | \vec \sigma \, | \, 0 \,
\rangle\, = -\hat e_+ / \sqrt{2} $.
With the substitutions 
$\vec Q = (\vec a
- \vec c \, ) = (\vec b + \vec {\bar c }\, )$ 
and $\vec \Sigma = (\vec a + \vec c \, )/2$,
$h_{fi}$
becomes
\begin{displaymath}
h_{fi}(d_{1q}) =
\underbrace{
\Big(-1\Big)
}_{signature}
\underbrace{
\Big(+{2^2\over 3^{3/2}}\Big)
}_{color}
\underbrace{
\Big(+{1\over 2^{1/2}}\Big)
}_{flavor}  \hskip 6cm
\end{displaymath}
\begin{equation}
\cdot \int \!\! \int {d^{\, 3}Q \, d^{\, 3}\Sigma \over (2\pi)^3 }  \
{
\langle   \vec \sigma \,   \rangle_{q\bar q} 
\cdot \vec Q
\over 2m_q } 
\ {\cal K} ( Q  ) \;
\phi_\rho( 2\vec \Sigma + \vec Q \, )
\;
\phi^*_{\pi^+}( 2\vec \Sigma + \vec Q  + \vec B \, )
\;
\phi^*_{\pi^o}( 2\vec \Sigma - \vec Q  + \vec B \, )
\end{equation}
Without additional information about the spatial wavefunctions this is
the final result. To proceed we assume standard quark model
Gaussian wavefunctions (A13) for the mesons.
The $d^{\, 3}\Sigma$ integral can then be carried out, which gives
\begin{displaymath}
h_{fi}(d_{1q}) =
-2^{-1} 3^{-3} \pi^{-15/4} m_q^{-1}  \beta^{-3/2}  \;
e^{-x^2 / 16} \; \hskip 5cm
\end{displaymath}
\begin{equation}
\hskip 5cm \cdot
\underbrace{
\int d^{\, 3}Q   \
\langle   \vec \sigma \,   \rangle_{q\bar q} 
\cdot \vec Q
\ {\cal K} ( Q \, )
\;
\exp\bigg\{ -{1\over 3 \beta^2}(\vec Q -{1\over 4}\vec B )^2\bigg\}
}_{I}
\end{equation}
where we have again abbreviated $P/\beta = x$.

We may evaluate this intermediate integral $I$ by returning to coordinate
space using (37) and integrating over $d^{\, 3}Q$, which gives
\begin{equation}
I = 3^{3/2}\pi^{3/2}\beta^3 \int  d^{\, 3} r \,
\langle \vec \sigma \rangle_{q\bar q} \cdot
({1\over 4}\vec B + {3i\over 2}\beta^2\vec r \, ) \ K(r) \
e^{\; {i\over 4} \vec B \cdot \vec r }\;  e^{- 3 \beta^2 r^2 / 4 } \ .
\end{equation}
For $\langle \vec \sigma \rangle_{q\bar q}$ in this reaction we have
\begin{equation}
\langle \vec \sigma \rangle_{q\bar q} \cdot \vec r = -\sqrt{ {2\pi \over  3 }} \,
r \, Y_{11}(\Omega) \ ,
\end{equation}
and the angular integrals can then be carried out using
a spherical harmonic expansion
of the plane wave in (C4). This gives
\begin{equation}
I = - 2^{5/2} 3\pi^3\beta^3 \; Y_{11}(\Omega_B)\; \int_0^\infty r^2\, dr
\ K(r) \;
e^{- 3 \beta^2 r^2 / 4 }
\bigg[
{\; 1\over 4}P\, j_0(Pr/4) - {3\over 2}\beta^2 r \, j_1 ( Pr/4)
\bigg] \ .
\end{equation}
Since the kernel is simply $K(r) = \alpha_s/r$ in this case, the
radial integrals can be evaluated in terms of confluent hypergeometric
functions using the general formula
\begin{equation}
\int_0^\infty  dr \, r^n \, j_\ell (ar) \; e^{-br^2} =
\sqrt{\pi}
\; {a^\ell \over b^\phi}
\;
{\Gamma(\phi) \over
2^{\ell+2} \Gamma(\ell + 3/2) }
\;
{}_1{\rm F}_1\Big({\ell - n\over 2}+1 ; \ell + {3\over 2} ; {a^2\over 4 b } \Big)\,
e^{-a^2/4b}  
\end{equation}
where $\phi = ( n+\ell+1)/2$.
This result suffices for all the overlap integrals we encounter
in JKJ decay models with SHO wavefunctions and power-law kernels.

Applied to the Coulomb case, our result for the integral $I$ is
\begin{equation}
I = -2^{3/2}\pi^3\alpha_s\beta^2 \, 
\, x
\bigg[
{}_1{\rm F}_1\Big({1\over 2};{3\over 2};\xi  \Big)
- {2\over 3}
{}_1{\rm F}_1\Big({1\over 2};{5\over 2};\xi  \Big)
\bigg] \, e^{-x^2/48} 
Y_{11}(\Omega_B) 
\ ,
\end{equation}
where 
$\xi = x^2 /  48 $.
Using this integral
we obtain our final result for diagram $d_{1q}$,
\begin{equation}
h_{fi}(d_{1q}) = {2^{1/2}\over 3^3} \pi^{-3/4}   {\alpha_s \over m_q}
\beta^{1/2} x\; \bigg[
{}_1{\rm F}_1\Big({1\over 2};{3\over 2};\xi \Big)
- {2\over 3}
{}_1{\rm F}_1\Big({1\over 2};{5\over 2};\xi  \Big)
\bigg]
\ e^{-x^2/12}
Y_{11}(\Omega_B)
\ .
\end{equation}
On multiplying by 4 for the four equivalent diagrams,
we obtain the full result for $h_{fi}$ for $\rho^+(+\hat z)\to \pi^+\pi^o$.
We can abstract the j$^0$Kj$^0$ decay amplitude ${\cal M}_{LS}$ in (6) from this;
removing the $\rho^+\to \pi^+\pi^o$ flavor factor
of $1/\sqrt{2}$, we have

\begin{equation}
{\cal M}_{10}^{ ( ^3{\rm S}_1 \to ^1{\rm S}_0 + ^1{\rm S}_0) } 
= +{2^3\over 3^3} \pi^{-3/4}   {\alpha_s \over m_q}
\beta^{1/2} x\; \bigg[
{}_1{\rm F}_1\Big({1\over 2};{3\over 2};\xi \Big)
- {2\over 3}
{}_1{\rm F}_1\Big({1\over 2};{5\over 2};\xi  \Big)
\bigg]
\ e^{-x^2/12}
\end{equation}
which is equivalent to the result quoted in (41). Using (6), we find
that the total $\rho\to\pi\pi$ decay rate due to color
Coulomb OGE (treated in isolation) is
\begin{equation}
\Gamma_{\rho\to\pi\pi}^{ {\rm j}^0{\rm Kj}^0 }
= \pi^{-1/2} \Big( {2^6\over 3^6} \Big)
\,  \alpha_s^2 \, \Big( {\beta \over m_q } \Big)^2
{E_\pi^2\over M_\rho }
x^3\;
\bigg[
{}_1{\rm F}_1\Big({1\over 2};{3\over 2};\xi \Big)
- {2\over 3}
{}_1{\rm F}_1\Big({1\over 2};{5\over 2};\xi \Big)
\bigg]^2 \,
e^{-x^2 / 6 } \ .
\end{equation}

Alternatively, for the specific case of a Coulomb kernel we can evaluate the
momentum space integral (C3) directly. The required integrals and 
others needed to evaluate decay amplitudes for
P-wave quarkonia are given below.

\begin{eqnarray}
&&\int  d^{\, 3} Q \; \Big( 1 / \vec Q^2\Big)  \; e^{-a(\vec Q - \vec Q_0 )^2} 
 = 
{\pi^{3/2}\over   a^{1/2}} \Bigg\{ 
\, 2 \, {}_1{\rm F}_1\Big({1\over 2},{3\over 2}, a\vec Q_0^2 \Big)
\Bigg\}
\; e^{-a\vec Q_0^2} \ ,
\\
&& \nonumber
\\
&&\int  d^{\, 3} Q \; \Big( Q_i /  \vec Q^2\Big) \; e^{-a(\vec Q - \vec Q_0 )^2} 
 = 
{\pi^{3/2} \over  a^{1/2} }\,  
\Bigg\{ {2\over 3}\,  
 Q_{0i}\; {}_1{\rm F}_1\Big({3\over 2}, {5\over 2}, a\vec Q_0^2 \Big)
\Bigg\} \; e^{-a\vec Q_0^2} \ ,
\\
&& \nonumber
\\
&&\int  d^{\, 3} Q \; \Big( Q_iQ_j / \vec Q^2 \Big) \; e^{-a(\vec Q - \vec Q_0 )^2} 
 = 
\nonumber\\
&& \hskip 1cm {\pi^{3/2}\over    a^{1/2} } 
\bigg\{
{1\over 3a}\, \delta_{ij} \; {}_1{\rm F}_1\Big({3\over 2}, {5\over 2}, a\vec Q_0^2 \Big)  
+ 
{2\over 5}  
\, Q_{0i} Q_{0j}\; {}_1{\rm F}_1\Big({5\over 2}, {7\over 2}, a\vec Q_0^2 \Big)
\bigg\}
\; e^{-a\vec Q_0^2} 
\ ,
\\
&& \nonumber
\\
&& \int d^{\, 3} Q \; \Big( Q_iQ_j / \vec Q^4 \Big) \; e^{-a(\vec Q - \vec Q_0 )^2} 
 = 
\nonumber\\
&& \hskip 1cm { \pi^{3/2}\over  a^{1/2} } 
\bigg\{
{2\over 3} \, \delta_{ij} \; {}_1{\rm F}_1\Big({1\over 2}, {5\over 2}, a\vec Q_0^2 \Big)
+{4\over 15}  a  
\, Q_{0i} Q_{0j}\; {}_1{\rm F}_1\Big({3\over 2}, {7\over 2}, a\vec Q_0^2 \Big)
\bigg\}
\; e^{-a\vec Q_0^2} 
\ ,
\\
&& \nonumber
\\
&& \int d^{\, 3} Q \; \Big( Q_iQ_jQ_k/ {\vec Q^4} \Big) \; 
e^{-a(\vec Q - \vec Q_0 )^2} 
 = 
\nonumber\\
&& \hskip 1cm {\pi^{3/2}\over  a^{1/2} }  
\bigg\{
{2\over 15} \, (\delta_{ij}Q_{0k} + \delta_{jk}Q_{0i} +\delta_{ki}Q_{0j} ) 
\; {}_1{\rm F}_1\Big({3\over 2}, {7\over 2}, a\vec Q_0^2 \Big)
\nonumber\\
&& \hskip 6cm +{4\over 35}  a  
\, Q_{0i} Q_{0j} Q_{0k}\; {}_1{\rm F}_1\Big({5\over 2}, {9\over 2}, a\vec Q_0^2 \Big)
\bigg\}
\; e^{-a\vec Q_0^2} 
\ .
\end{eqnarray}
This approach applied to $\rho\to\pi\pi$
leads to the form
\begin{equation}
\Gamma_{\rho\to\pi\pi}^{ {\rm j}^0{\rm Kj}^0 }
= \pi^{-1/2} \Big( {2^6\over 3^8} \Big)
\,  \alpha_s^2 \, \Big( {\beta \over m_q } \Big)^2
{E_\pi^2\over M_\rho }
x^3
{}_1{\rm F}_1\Big({3\over 2};{5\over 2};\xi \Big)^2 \, 
e^{-x^2 / 6 } 
\end{equation}
which is equivalent to (C12).

Relations between confluent hypergeometric
functions with different indices allow these results to be written in 
various forms. We typically express our final results for decay amplitudes
as linear combinations of
confluent hypergeometric functions with constant coefficients and a common
first index $a$, times a centrifical factor of $x^{L_{BC}}$.
These rearrangements are straightforward using the recurrence relations
\begin{equation}
{}_1{\rm F}_1\Big(a; c; x \Big)
=
{(c-1) \over (a-1)}
{}_1{\rm F}_1\Big(a-1; c-1; x \Big)
+ {(a-c) \over (a-1)}
{}_1{\rm F}_1\Big(a-1; c; x \Big)
\end{equation}
and
\begin{equation}
x {}_1{\rm F}_1\Big(a; c; x \Big)
=
{(c-1)(c-2) \over (a-1)}
\bigg\{
{}_1{\rm F}_1\Big(a-1; c-2; x \Big)
- {}_1{\rm F}_1\Big(a-1; c-1; x \Big)
\bigg\}  \ .
\end{equation}

The decay rates due to transverse OGE and
the confining interaction may be derived using the same
techniques, with the minor complication that transverse OGE also
has initial-line spin-flip contributions. These results are quoted in
Section II.D.


\begin{references}
\bibitem{gammanote} Several conventions for $\gamma$ are
currently in use, which typically differ in the overall normalization
of equation (1).
\end{references}
\end{document}